# Uncertainty Quantification of Large-Eddy Simulation Results of Riverine Flows: A Field and Numerical Study

K. Flora[1], and A. Khosronejad[1]

[1]Civil Engineering Department, Stony Brook University, Stony Brook, NY 11794

Corresponding author: Ali Khosronejad (ali.khosronejad@stonybrook.edu) Abstract

**Abstract**

We present large-eddy simulations (LESs) of riverine flow in a study reach in the Sacramento River, California. The riverbed bathymetry was surveyed in high-resolution using a multibeam echosounder to construct the computational model of the study area, while the topographies were defined using aerial photographs taken by an Unmanned Aircraft System (UAS). In a series of field campaigns, we measured the flow field of the river using the acoustic Doppler current profiler (ADCP) and estimated using large-scale particle velocimetry of the videos taken during the operation UAS. We used the measured data of the river flow field to evaluate the accuracy of the LES-computed hydrodynamics. The propagation of uncertainties in the LES results due to the variations in the effective roughness height of the riverbed and the inflow discharge of the river was studied using uncertainty quantification (UQ) analyses. The polynomial chaos expansion (PCE) method was used to develop a surrogate model, which was randomly sampled sufficiently by the Monte Carlo Sampling (MCS) method to generate confidence intervals for the LES-computed velocity field. Also, Sobol indices derived from the PCE coefficients were calculated to help understand the relative influence of different input parameters on the global uncertainty of the results. The UQ analysis showed that uncertainties of LES results in the shallow near bank regions of the river were mainly related to the roughness, while the variation of inflow discharge leads to uncertainty in the LES results throughout the river, indiscriminately.

## 1 Introduction

The use of computational fluid dynamic (CFD) models has become more prevalent in hydraulic modeling of natural systems in recent years due to improvements in high-performance computing capabilities. Advanced numerical models can provide meaningful insights into the three-dimensional (3D) flow structures naturally occurring in riverine environments. For example, 3D computational models can help to understand better the dynamics of the secondary flow at meander bends, and vortex shedding processes from man-made obstructions such as bridge foundations and spur dikes (Kang et al., 2020, 2021; Khosronejad, Flora, & Kang, 2020; Khosronejad, Flora, Zhang, et al., 2020; Khosronejad & Sotiropoulos, 2017). The early use of CFD models for the study of rivers was limited by computing power to relatively coarse computational meshes and employed time-averaged Reynolds Averaged Navier-Stokes (RANS) turbulence closure models (Fischer-Antze et al., 2008; Liang & Fotis, 2005; Liu & García, 2008; Rüther et al., 2010; Wilson et al., 2003). Recently, more sophisticated, high-fidelity models using detached eddy simulation (DES) and large eddy simulation (LES) have provided the capability to capture even greater spatial and temporal details of the flow (Constantinescu et al., 2011; Kara et al., 2015; Khosronejad,

Kozarek, Diplas, et al., 2015; Khosronejad, Kang, et al., 2019). For instance, LES models for full-scale rivers like the Mississippi River in Minnesota and the Feather River in California have recently been modeled using LES (Khosronejad, Le et al., 2016; Khosronejad, Flora, & Kang, 2020). Despite these massive improvements in the resolution of the hydrodynamic in rivers, little attention has been paid to quantifying the results' relative uncertainty. Highly detailed flow structures may have resulted from considerable uncertainty in modeling parameters and methods, leading to an excessive level of confidence being placed in modeling results. Furthermore, the precision offered by highly detailed and impressive results derived from sophisticated numerical models that were processed by overwhelming computing power may still lead to misleading or, even worse, incorrect interpretation.

Uncertainty in numerical modeling arises from many factors, which can be broadly divided into five primary types: (1) modeling uncertainty, (2) discretization and truncation errors, (3) convergence errors, and (4) computer round-off errors (Sadrehaghighi, 2020). Some of these sources of error can be estimated or minimized through increased computational effort, such as using higher-order numerical schemes, smaller temporal and spatial steps, and specifying reduced tolerance for reaching convergence (Celik et al., 2008). However, the uncertainty associated with modeling is often due to model simplifications, limited knowledge of model parameters, boundary conditions, or accurate calibration data. These types of sources of uncertainty cannot be eliminated or easily reduced, which makes it necessary to quantifying these types of errors to enable a more comprehensive understanding of the model results. In particular, hydrodynamic modeling of natural waterways has many sources of uncertainty. This is due to the spatial nature of the domain, which is typically large, highly complex, and whose boundaries are frequently changing. In a typical river, heterogeneity of the channel bed materials, variations in density and types of vegetation, and complex shapes on the bed, including bedforms and pools, make it impossible to capture this degree of variation fully and requires many modeling simplifications (Casas et al., 2010; Flora & Khosronejad, 2021; Lawless & Robert, 2001; Warmink et al., 2011).

In addition to topography, the inflow boundary conditions could be complicated by hydrological ambiguity and the error associated with the flow field measurements. Being generated by rare meteorological events, a sufficiently large number of historical flow rates are often lacking at the model location, leading to the statistical uncertainty for desired flows of interest. Additionally, hydrologic uncertainty remains regarding the recorded flow rate data, which are typically derived from rating curves with inherent errors in both the method and the associated discharge measurements (Di Baldassarre & Montanari, 2009; Pelletier, 1988). Nowadays, Acoustic Doppler Current Profilers (ADCPs) are commonly used for estimating discharges in natural streams because of their relative speed and accuracy. However, ADCPs are known to have many potential sources of error, including operator error, programming errors, instrument errors, and extrapolation errors near the bed, water surface, and channel banks (Mueller & Wagner, 2009; Muste et al., 2010; Muste, Yu, & Spasojevic, 2004; Schmalz et al., 2012). Additionally, when used for calibration of numerical models, ADCP data can suffer from the practical difficulty of obtaining enough passes over the same transect line to provide a reliable time-averaged velocity from the instantaneous data it measures. Nevertheless, with proper operation and post-processing (Muste, Yu, Pratt, et al., 2004), it is possible to determine accurate average turbulence intensities and time-averaged velocity field by equally sampling fluctuations higher and lower fluctuations than the mean value (Fernández-González et al., 2017; Muste, Yu, & Spasojevic, 2004). Uncertainty in high-fidelity modeling associated with ADCP measurements used for inflow flow rates was explored in this study.

Another field measurement for riverine models which can be subject to error is inaccurate or incomplete mapping of the bathymetry of the channel (Casas et al., 2006; Legleiter et al., 2011). In a case study, Pasternack et al. (Pasternack et al., 2006) estimated that 21 percent of the predicted depth error in a two-dimensional (2D) model could be attributed to the errors in topographic resolution. Moreover, in 3D modeling, accurate bathymetry data is critical for determining not only mean flow characteristics such as depth and mean velocity values but turbulent structures and secondary velocity components, as well (Keylock et al., 2012; Lane et al., 1999). The uncertainty associated with delineation of the channel bed surface can result from several sources of measurement error such as faulty echosounder readings, spatial inaccuracies owing to poor quality GPS positions, and unaccounted rotations of the boat and sonar transducer. Besides, the low spatial resolution of the sonar readings, caused by low sampling rates or overly wide spacing between adjacent vessel passes, can result in missing essential features in the bathymetry and produce overly simplified channel surfaces. Many of these errors can be reduced or eliminated using (1) more precise equipment such as augmenting GPS with an inertial navigation unit (INU) for improved spatial accuracy and (2) a multibeam echosounder (MBES) when possible, instead of a single-beam echosounder. A comparison of LES results from MBES and single-beam sonar data showed improved bathymetric resolution reduced near-bed velocities and shear stresses, but increased the number of turbulent structures in areas with highly irregular topography (Flora & Khosronejad, 2021).

Resistance in a channel can be due to many sources, including instream obstacles such as vegetation and bridge piers, bank irregularities, meander bends, bedforms, and roughness due to individual grain particles (Recking et al., 2008; Robert, 2003). While the larger features may be captured in bathymetric/topographic survey data and resolved directly in the discretization of a high-fidelity model, the surface roughness of the individual grains cannot be practically resolved when modeling a natural river. Instead, wall models are employed to account for this type of resistance. Wall models capture different degrees of surface roughness or grain sizes by assigning an equivalent roughness height value, $k_s$, which is proportional to a characteristic grain size. Unfortunately, there is no general consensus among researchers for the constant of proportionality or which specific grain size among the distribution of grain sizes one should use (Camenen et al., 2006). Chanson (2004) has provided a summary of equivalent roughness heights of several past researchers regarding flow resistance, which ranged between $1.5d_{50}$ by Sumer (1996) in cases of sheet flow to $3d_{90}$ by van Rijn (1984) in bedload transport of experimental and field data. Other researchers have argued that $k_s$ is not only a function of grain size but is a function of the bed load transport (Camenen & Larson, 2013; Sumer et al., 1996; Yalin, 1992) and the dimensionless settling velocity rate, as well (Camenen et al., 2006). Regardless, the general lack of consensus creates uncertainty in what value of $k_s$ is appropriate for use in wall models. This study seeks to investigate how the grain roughness uncertainty impacts the hydrodynamics simulation results of a field-scale river.

Hence, the objective of this study is to (1) collect bathymetric and flow measurement data for a case study in a reach of the Sacramento River in California, (2) conduct high-fidelity large-eddy simulations (LESs) of the river flow, and (3) quantify the uncertainty in the hydrodynamic results of the river flow for two input modeling parameters: discharge and the equivalent roughness height. Specifically, the inflow rate will be varied based on the measurement uncertainty identified from the ADCP measurements at multiple transects in the river reach. Also, a range of equivalent roughness height parameters will be used in the wall model to determine how changing the characteristic grain size will alter the flow velocities of this natural river. To quantify the combined uncertainty of the LES results, polynomial chaos expansion (PCE) and Monte Carlo (MC) simulations will be used to provide approximate velocities at select transects over a continuous range of input parameters to generate a 95 percent confidence levels for the computed velocity fields. Sobol indices, also known as "variance-based sensitivity analysis," which are used as a form of global sensitivity analysis, were determined to estimate the relative impact each of these two parameters has on the hydrodynamics simulation results. Finally, the computed depth-averaged velocity magnitudes along ADCP transect lines and velocity profiles at select locations were compared against the measured flow field data.

The governing equations and the wall model used in the numerical model are introduced in Section 2. Next, the case study and the field data collection for the river bathymetry and flow field measurement are discussed in Section 3. Subsequently, the computational details, boundary conditions, and the range of input parameters for the LES are described in Section 4. Afterward, the uncertainty quantification procedures are described in Section 5, followed by presenting and discussing the LES results and UQ analysis of the computed flow field in Section 6. Finally, conclusions concerning the importance of assessing uncertainty in numerical modeling of natural rivers are presented in Section 7.

## 2 Numerical Framework

To simulate instantaneous, 3D, incompressible, turbulent flow for our case study, we use our in-house, open-source CFD code, the so-called Virtual Flow Simulator (VFS-Geophysics) model. The successful use of VFS-Geophysics for simulating flow in natural waterways has been well documented (Giang & Hong, 2019; Kang & Sotiropoulos, 2012; Khosronejad, Le, et al., 2016; Khosronejad, Kang, et al., 2019; Khosronejad, Flora, & Kang, 2020). For implementing LES in VFS-Geophysics model, the curvilinear immersed boundary (CURVIB) method described by Ge and Sotiropoulos (Ge & Sotiropoulos, 2007) is used to resolve the arbitrarily complex geometric shapes associated with a natural river bathymetry such as channel bedforms, vegetation, bridge piers, and irregularly shaped banks. In the context of the Immersed Boundary Method (IBM), a structured background mesh is used for the flow domain. The solid boundaries of objects such as the channel banks and channel bed are treated as unstructured triangular meshes, which are superimposed inside the flow domain (Fig. 1). The IBM classifies each node inside the flow domain as either a solid, fluid, or immersed boundary (IB) node -- as depicted in Fig. 2. As seen, the nodes shown in the slice in Fig. 2a are designated in Fig 2b as either (1) solid nodes (red region) which are removed from the computations, (2) fluid nodes (blue region) which reside outside of the solid boundary where the governing equations for the flow computations are applied, or (3) the IB nodes (white circles within the gray region) adjacent to the boundary where the velocities are reconstructed using a wall model.

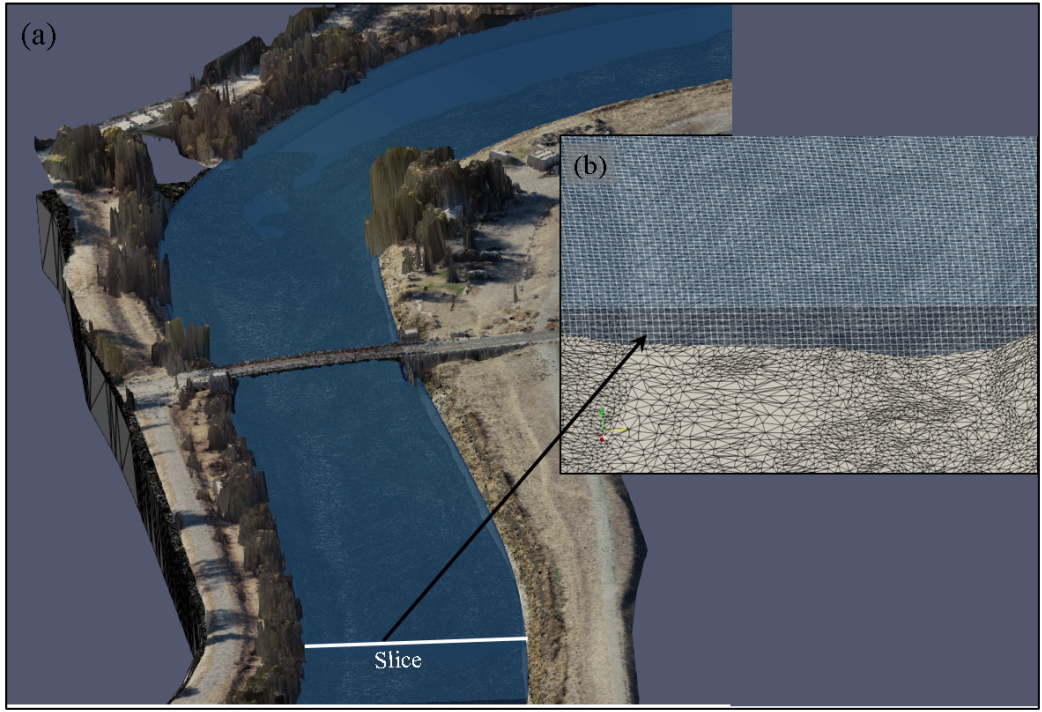

**Figure 1.** Schematic of the flow domain. The slice outlined in (a) is shown enlarged in (b) with the white lines representing the structured grid system of the flow domain, and the black triangles on the bed are the unstructured mesh of the IB.

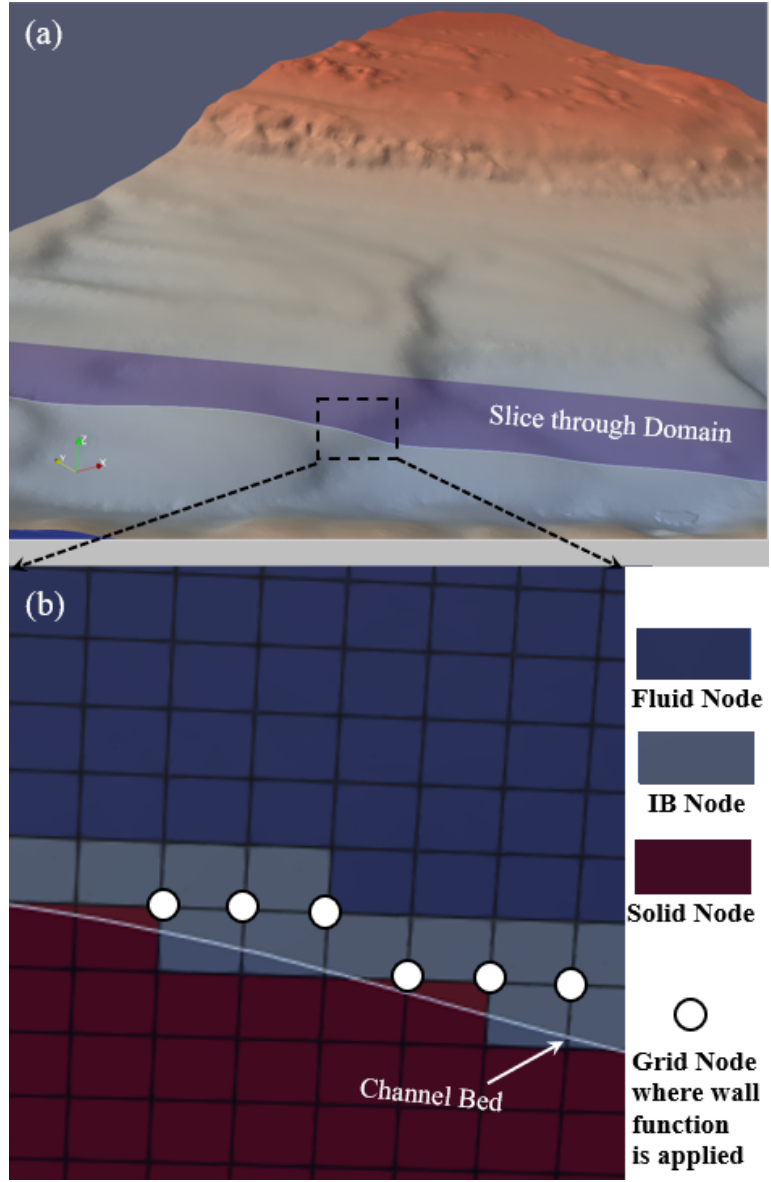

**Figure 2.** (a) Schematic of a slice taken through channel bed. (b) is an enlarged view of the area near the riverbed. (b) shows the classification of the background mesh into fluid, solid, and IB nodes and the location of the nodes adjacent to the wall boundary, where the wall models are applied.

In the VFS-Geophysics model, the implementation of LES is fully described by Kang et al. (Kang et al., 2011) and uses the following algorithm from the Smagorinsky sub-grid scale (SGS) model (Smagorinsky, 1963);

$$\tau_{ij} - \frac{1}{3}\tau_{kk}\delta_{ij} = -2\mu_t \overline{S}_{ij} \qquad (1)$$

where μt is the eddy viscosity, and Sij is the large-scale strain-rate tensor. The eddy viscosity, $\mu_t$, is defined as:

$$\mu_t = C_s \Delta^2 |\overline{S}| \qquad (2)$$

where $C_s$ is the Smagorinsky constant, $\Delta$ is the filter width determined by the grid resolution, and $|\bar{S}|$ is the magnitude of the strain-rate tensor obtained from $|\bar{S}| = \sqrt{2\bar{S}_{ij}\bar{S}_{ij}}$.

In the flow domain, Navier-Stokes equations for incompressible flow are solved in the hydrodynamic model. The conservation of mass for incompressible flow is expressed as (Ferziger et al., 2020):

$$\frac{\partial u_j}{\partial x_j} = 0 \tag{3}$$

where $u_j$ represents the spatially-filtered velocity components and the conservation of momentum reads as (Khosronejad & Sotiropoulos, 2014):

$$\frac{\partial u_i}{\partial t} + u_j \frac{\partial u_i}{\partial x_j} = -\frac{1}{\rho}\frac{\partial P}{\partial x_i} + \nu \nabla^2 u_i + g_i \tag{4}$$

where $\rho$ is the density of water, P is the pressure, $\nu$ is the kinematic viscosity of water, and g is the gravitational body force.

A wall model is used to solve indirectly for the velocity at the first grid point off the solid boundaries (Fig. 2b) for cases when the viscous sublayer itself cannot be resolved directly, such as in a large flow domain like a natural river and where the individual grains are too small to be resolved. The wall modeling approach assumes that the first grid points off the wall reside within the log layer and reads as follows (Khosronejad et al., 2011):

$$\frac{u}{u_*} = \begin{cases} y^+ & y^+ \leq 11.53 \\ \frac{1}{\kappa} ln(Ey^+) & y^+ > 11.53 \end{cases} \tag{5}$$

where $u$ is the local velocity magnitude at the node located a distance, $y$, from the wall, $u_* = \sqrt{\frac{\tau}{\rho}}$ is the shear velocity (Wilcox, 1993), $\kappa$ is von Karaman's constant ($\approx 0.41$), and $y^+$ is the distance of the first node off the wall in wall unit ($= \frac{yu_*}{\nu}$), and E is the roughness parameter, defined as follows:

$$E = exp[K(B - \Delta B)] \tag{6}$$

where

$$\Delta B = \begin{cases} 0 & k_s^+ < 2.25 \\ \left[B - 8.5 + \frac{1}{\kappa}ln(k_s^+)\right] sin[0.4258(ln(k_s^+) - 0.811)] & 2.25 < k_s^+ < 90 \\ B - 8.5 + \frac{1}{\kappa}ln(k_s^+) & k_s^+ \geq 90 \end{cases} \tag{7}$$

where B is a constant (=5.2) and $k_s^+$ is the roughness Reynolds number, $\frac{k_s u_*}{\nu}$, where $k_s$ is the equivalent roughness height. The three cases for $\Delta B$ in Eqn. 7 represent different effects on the flow caused by how far above the channel boundary the roughness elements extend, relative to the viscous sublayer, $\delta_v$, where viscous forces dominate. When $k_s^+ < 2.25$, the roughness elements are entirely within $\delta_v$, where the turbulence generated by roughness elements themselves are fully dampened out by the viscous forces and do not contribute to the overall channel resistance, which is referred to as hydraulically smooth flow. However, in natural rivers, this condition typically does not exist since roughness elements generally extend beyond $\sim 5\delta_v$ (Robert, 2003). For this study, $k_s$ will be varied to observe how changes in the equivalent roughness height in the wall model influence the hydrodynamics of the river.

## 3 Case Study: Sacramento River at Knights Landing

### 3.1 Site Description

The Sacramento River is a major river in Northern California with headwaters in the Klamath Mountains which flows over 700 km into the San Francisco Bay-Delta. Approximately 50 km upstream from the City of Sacramento, the Sacramento River makes a sharp bend -- as it passes through the town of Knights Landing, CA, which is the site for this case study (Fig. 3). The reach of interest for our simulations is approximately 610 m long, with a two-lane highway bridge on Route 113 crossing the river just downstream from the meander bend. The bridge has piers at each bank protected by timber fenders causing only a minor contraction of flow at the crossing (Fig. 4). The river in this location is confined by levees on each bank and has an average channel top width of about 70 m during our field data collection stage. During our field investigation, the mean flow depth in the channel was about 3.3 m with a deep section approximately 8.5 m deep along the outer bank located at the channel bend. A side tributary, called Sycamore Slough, enters the Sacramento River along the right bank near the midlength of the study area. Irrigation water flows into the slough through a gate located about 400 m upstream from the confluence on Sycamore Slough; however, this inflow is considered an insignificant addition to the main channel flow in the Sacramento River.

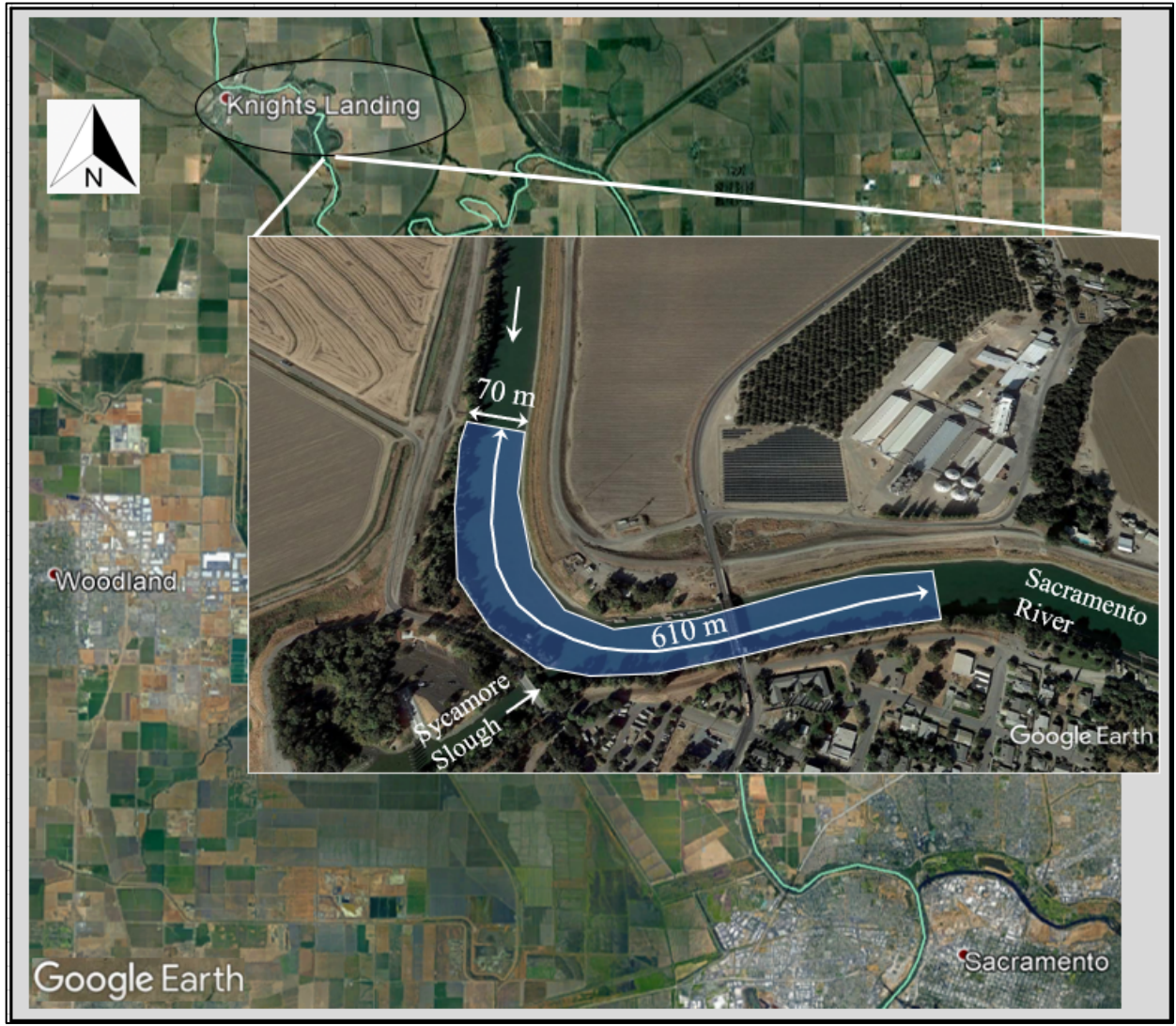

**Figure 3.** Site map of the study area in the Sacramento River in Knights Landing, California. Flow is from the North to the East.). (Earth data © 2020 Google.)

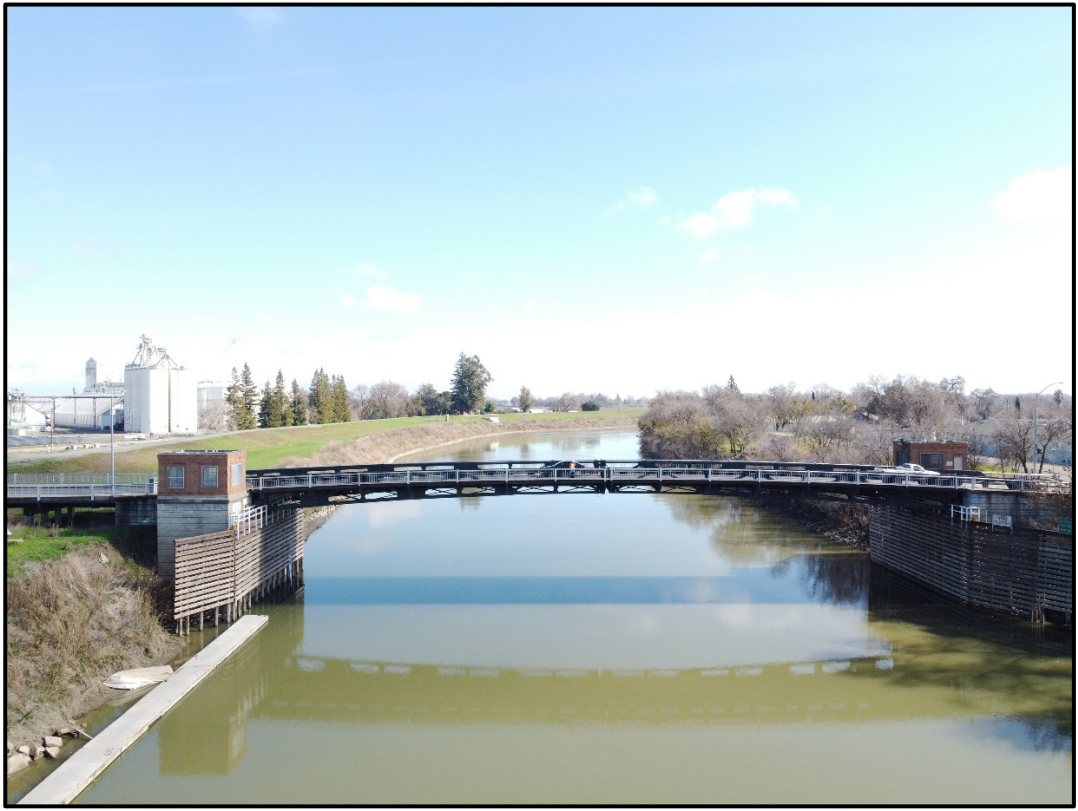

**Figure 4.** Looking downstream at the Highway 113 bridge crossing over the Sacramento River at Knights Landing, showing piers protected by timber fenders along each bank.

3.2 Field Measurements

In December 2020, we collected three types of field data: (1) Aerial photogrammetry and video imagery, (2) ADCP flow field, and (3) bathymetric survey. Herein, the measurement methodology, equipment used, and the results of each type of data collection are described briefly below. Videos and images were taken at the site using a DJI Phantom 4 unmanned aircraft system (UAS). Almost 700 4K still photos were taken from an altitude of about 60 m and used to generate topographic data. This imaging technique is based on the Structure from Motion (SfM) (Fonstad et al., 2013) methodology, which uses open-source software of OpenDroneMap (*Drone Mapping Software - OpenDroneMap*, n.d.). The elevation data above the water surface (Fig. 5) was tied into the 30 ground control points surveyed using a Trimble GeoExplorer 7x GPS unit. This unit allows for augmenting the bathymetric survey data to generate a complete digital map of the river within the study reach.

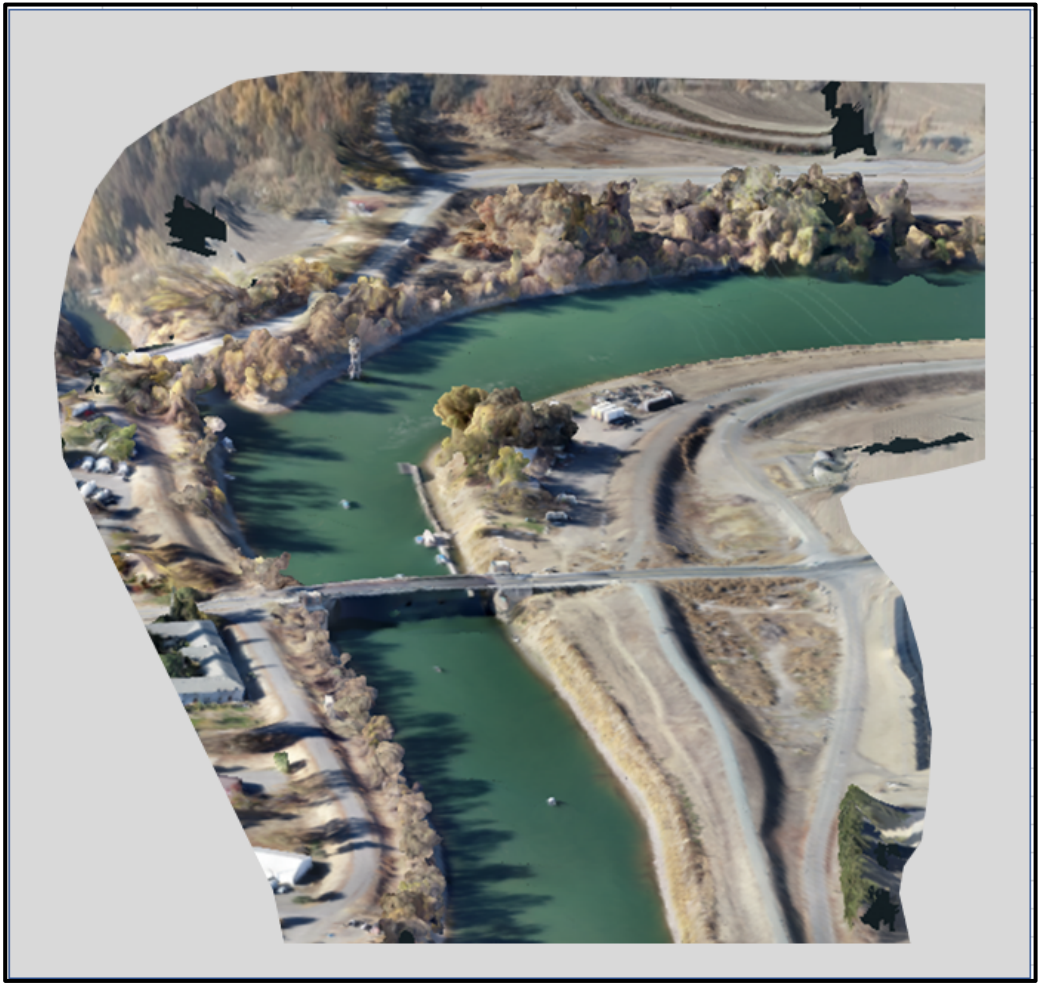

**Figure 5.** Textured model showing the measured above water topographic data, derived from aerial photography using SfM for Sacramento River at Knights Landing. Flow is from the top right to the bottom.

During the same day, over 20 aerial movies were taken from the field site using the same DJI Phantom 4 UAS. These animations were used to produce still images, which were then processed using Large-Scale Particle Image Velocimetry (LSPIV) to obtain the water surface velocity field of the river. LSPIV was first used by Fujita (1997) to monitor a flood event in Japan in 1993. Since then, others have successfully employed this method to estimate the water surface velocity field and turbulence of riverine environments (Fujita et al., 1998; Jin & Liao, 2019; Lewis & Rhoads, 2015; Sutarto, 2015). LSPIV is an extension of Particle Image Velocimetry (PIV), commonly used in laboratories where water is seeded with particles and tracked with high-speed cameras to study flow dynamics in flumes. With LSPIV, naturally or artificial seeded particles floating on the water surface are similarly tracked over multiple video frames to estimate the surface velocity magnitude and the direction of the flow. Herein, the videos were shot normal to the water surface from a stationary UAS for between 30 and 60 seconds and included at least two control targets with measured coordinates within the field of view. The control targets enable the scaling of the measurements during the post-processing of the data. The water surface flow characteristics were analyzed for the most promising videos based on the quantity of seeding using open source software, i.e., RIVeR (Patalano et al., 2017) using only natural seeding consisting primarily of bubbles and small vegetal debris on the water surface. This software is a graphical user interface for the MATLAB tool PIVlab designed specifically for LSPIV processing. As seen in Fig. 6, the seeding and lighting on the water surface varied significantly and strongly influenced the quality of the results. In Fig. 6(b), Regions 2 and 4 have minimal seeding and poor lighting conditions. Therefore, the LSPIV processing could not effectively resolve the velocity directly -- requiring the program to extensively interpolate these regions from the resolved velocities in Regions 1 and 3. Despite minimal seeding, Region 1 still provided reasonable velocity data, which compared favorably with the other velocity measurements (see Section 6). However, the excessive uncertainty in the LSPIV velocities proved too great to provide useful flow field data for this study. The computed depth-averaged results taken as 85 percent of the surface velocities are presented in Section 6.

On December 15, 2020, ADCP streamflow measurements were collected using a Teledyne RD Workhorse Rio Grande® 1200 kHz ADCP and their WinRiver® II data acquisition software. Spatial data for the survey was obtained using the Advanced Navigation Spatial Dual® Inertial Navigation Unit (INU). Based on the measured stage from LSPIV data collection of the previous week, the stage of the river was nearly identical. In total, 17 ADCP transects were made for the Sacramento River at 5 locations, with three locations being upstream from the highway bridge and two locations downstream -- see Fig. 7. Flow rates from each transect are shown in Table 1, with an average flow rate of 130.5 m3/s with a standard deviation of 4.7 $m^3$/s. Variability in the measurements could be primarily attributed to operator error and a minor inflow from the Sycamore Slough stream that was measured to be equal to 5.3 $m^3$/s. Each of the five transect locations was post-processed using the Velocity Mapping Toolbox (VMT) software available from the USGS (Parsons et al., 2013). Using VMT, multiple transects at each location were combined to provide a single set of velocity values along a spatially-averaged cross-section to provide a more time-averaged depiction of the flow than the instantaneous velocities measured during each transect (Fig. 8). The velocity values were also depth-averaged and used for comparison with the LES simulations.

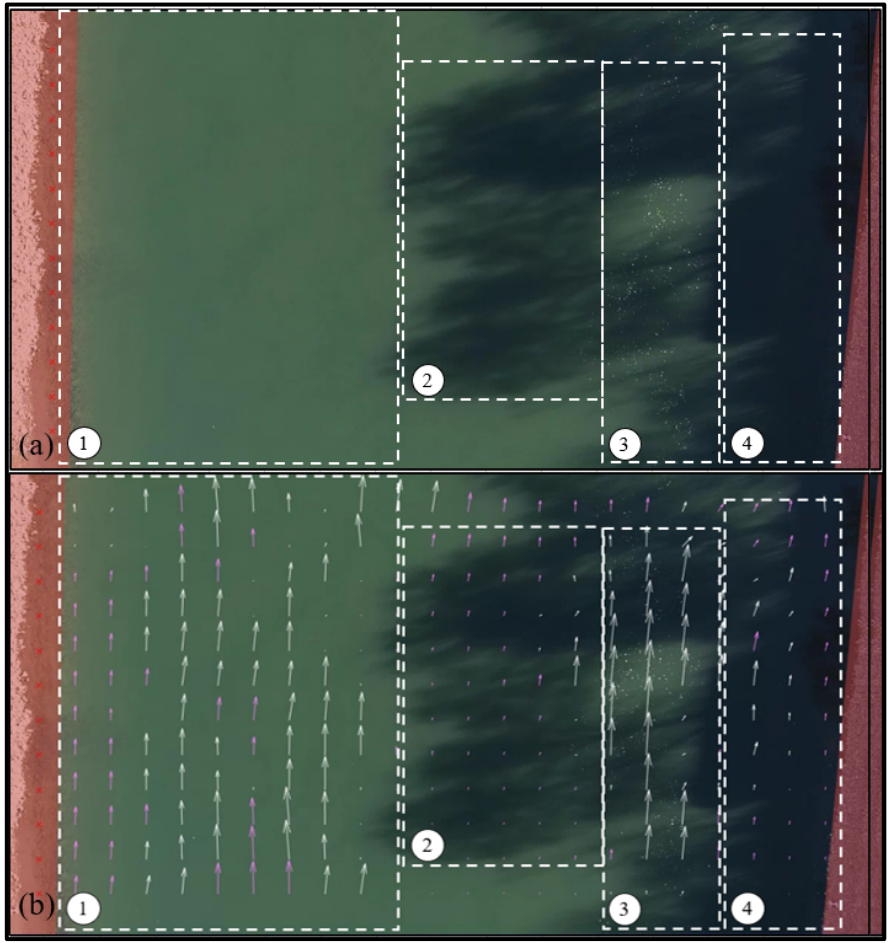

**Figure 6.** (a) An aerial image of the water surface immediately downstream from the highway bridge. Region 1 has bright sunlight but negligible seeding. Regions 2 and 4 are in the shadows with poor lighting and minimal seeding. Region 3 has mixed lighting and substantial seeding. (b) The white arrows are the computed velocity vectors, while the pink arrows are velocity vectors based on the interpolation.

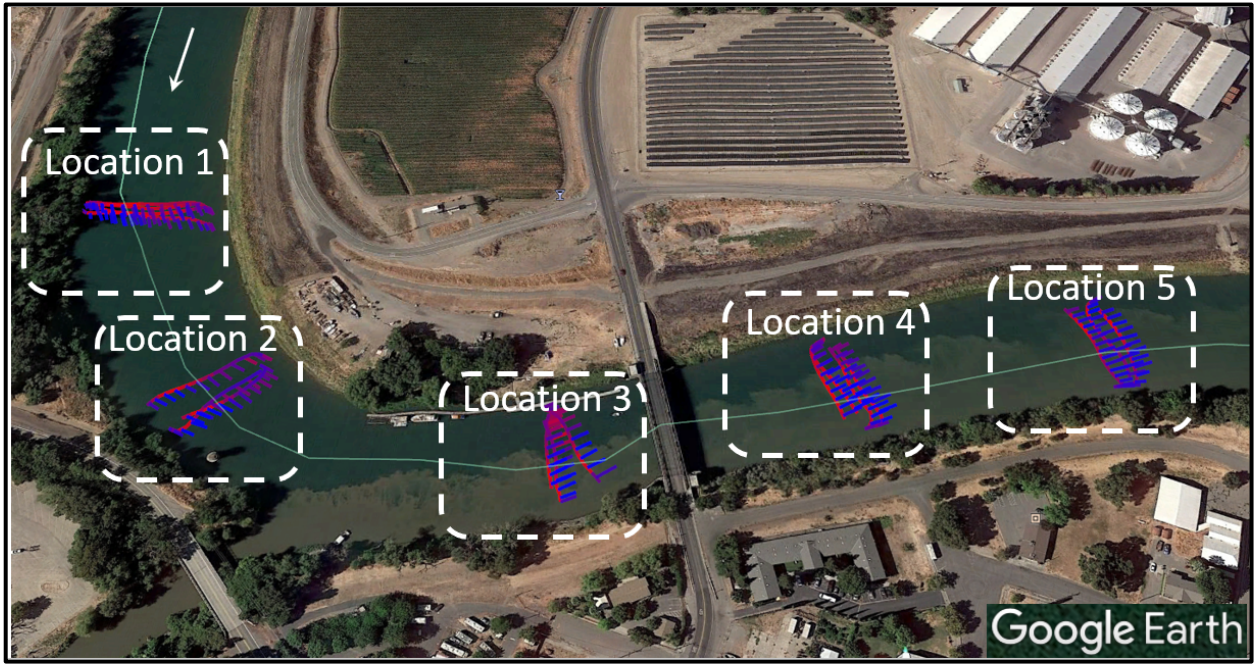

**Figure 7.** Locations of the ADCP measurements. The individual transects demonstrating the depth-averaged flow velocities are presented in blue.

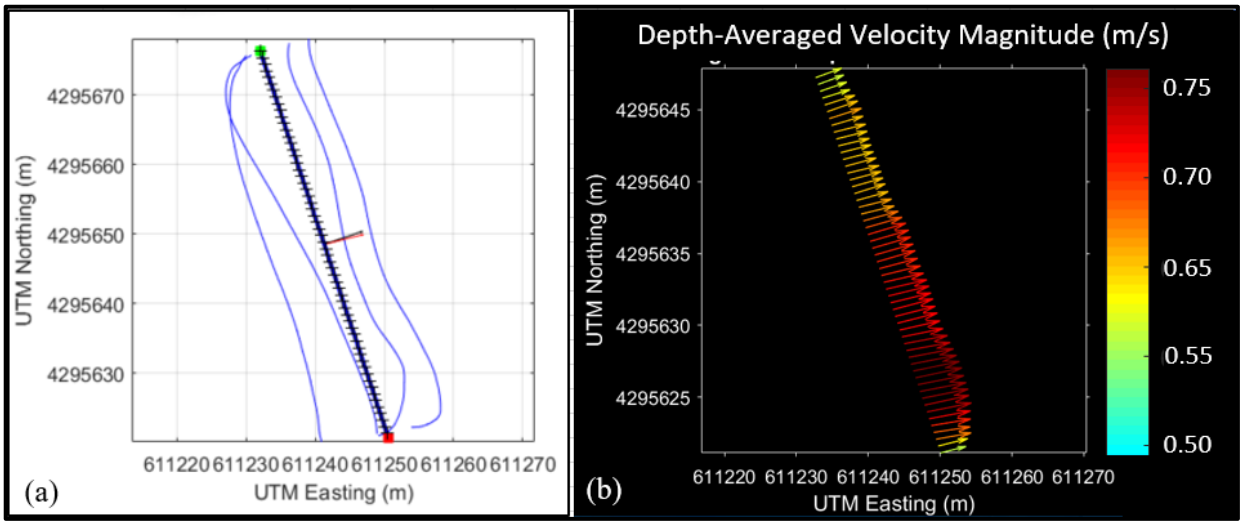

**Figure 8.** Data post-processing using VMT software. (a) Individual transect lines (blue) and smoothed averaged transect line (black) for Location 4. (b) depth-averaged velocities along the smoothed transect line.

On the same date as the ADCP measurements, a bathymetric survey of the channel bed was made using a Picotech PICOMB-120 MBES, which operated at 25 pulses per second. The PICOMB-120 has a 120-degree swath angle projecting 256 beams equally spaced at 0.47 degrees. Overlapping passes of the operating boat were used to measure the detailed bed bathymetry of the river. The boat movements and locations were measured using the Advanced Navigation Spatial Dual INU and post-processed for improved coordinates using Advanced Navigation's

KinematicaTM algorithms. The heading for the INU was provided using 2 Trimble Zephyr 2 antennas. Over 26 million bathymetric data points were collected and, subsequently, post-processed using the BeamworXTM Hydrographic Suite of software – see Fig. 9. As clearly seen, the survey data captured large bedforms that are 1 to 1.5 m high and present throughout the reach. Also, scour regions that are approximately 4 m deep can be seen along the outer bank of the meander bend, while about 2 m deep scour regions were detected near the bridge crossing. As shown in the deeper parts of the river (in blue), the thalweg of the river switches sides by crossing over from the right bank at the bridge crossing to the left bank near the downstream end of the study reach. This shift in thalweg is typical of a meandering river and influences the bathymetry and hydrodynamics, as will be seen in the LES results presented in Section 6.

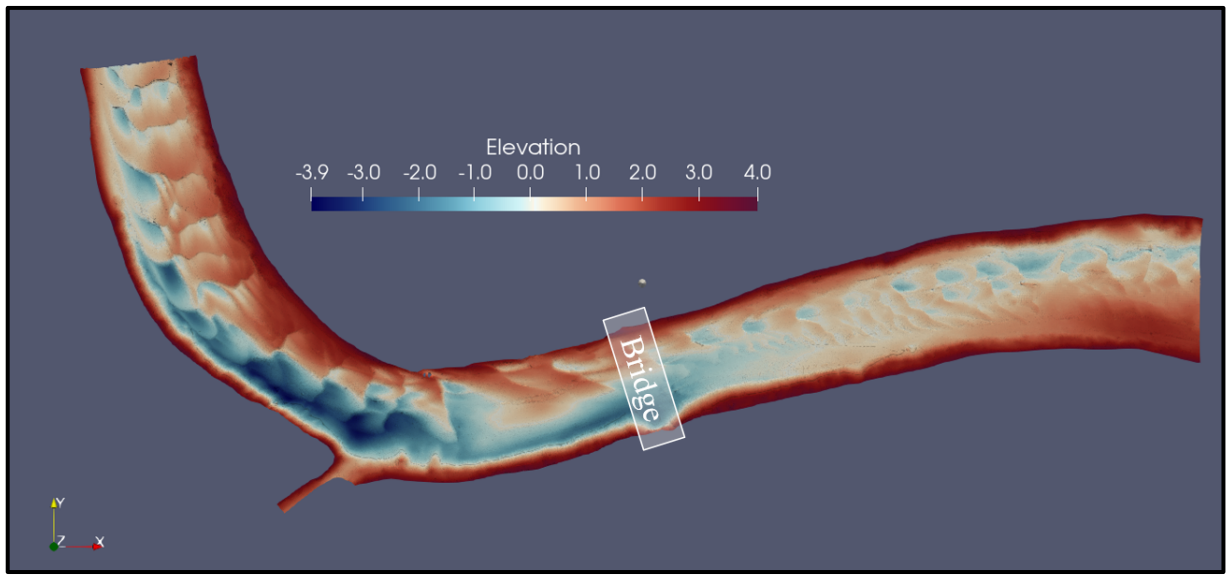

**Figure 9.** Bathymetric survey data for the Sacramento River at Knights Landing. Color maps show the relative bed elevation of the river in meters.

**Table 1.** Summary of ADCP Measurements. Locations are shown on Fig. 7.

| Location | Transect | $Q$ (m³/s) |
|---|---|---|
| 1 | 1 | 131.9 |
| | 2 | 122.4 |
| | 3 | 128.5 |
| | 4 | 130.7 |
| 2 | 5 | 124.5 |
| | 6 | 124.0 |
| | 7 | 122.9 |
| 3 | 8 | 133.9 |
| | 9 | 129.9 |
| | 10 | 136.8 |
| 4 | 11 | 132.8 |
| | 12 | 136.6 |
| | 13 | 134.5 |
| | 14 | 136.3 |
| 5 | 15 | 130.6 |
| | 16 | 132.6 |
| | 17 | 130.4 |

## 4 Computational Details

The LES of the river flow in the study area was carried out for a mean-flow depth of 3.3 m and mean-flow rate of 0.565 m3/s, which result in the Reynolds and Froude numbers of $1.86 \times 10^6$ and 0.11, respectively. The flow domain was discretized using a uniform grid node spacing in all directions equivalent to ~ 0.6 m. For this spacing, 201, 1201, and 21 nodes were placed in the spanwise, streamwise, and vertical directions, respectively, to create a flow domain with approximately 5 million computational grid nodes. About half of these nodes were classified as solid nodes and banked out of the computations. The timestep for all simulations was equivalent to 0.07s. Typical values for $y^+$ were greater than 1000. Given the Reynolds number of the river flow in this study, the employed grid resolution is understandingly too coarse to capture the viscous sublayer of the turbulent flow. However, since in such large-scale riverine flows, large bathymetry features and other wall-mounted structures are responsible for producing slowly evolving but very energetic coherent structures, which are responsible for producing most of the turbulence, we have been able to successfully resolve such flows using wall models (Flora et al., 2021; Khosronejad et al., 2011, 2012, 2013; Khosronejad, Diplas, Angelidis, et al., 2020; Khosronejad, Flora, & Kang, 2020; Khosronejad, Kozarek, Palmsten, et al., 2015). That is without needing to resolve the details of the near wall flow, provided that one uses grids fine enough to capture key geometrical features and resolve the slowly coherent flow structures they induce. This has been the basic premise for the success of LES linked with the immersed boundary method developed for and introduced in simulations of river hydrodynamics and morphodynamics (Khosronejad et al., 2013; Khosronejad, Flora, & Kang, 2020; Khosronejad, Flora, Zhang, et al., 2020; Khosronejad, Le, et al., 2016).

Based on the field observations of the water surface elevation during the bathymetric survey, there was minimal variation in the water surface throughout the reach. The water slope of the river was highly mild (0.005percent). For this reason (Khosronejad, Ghazian Arabi, et al., 2019), we employed a flat rigid-lid assumption at the free surface to facilitate the computations of the LESs using the location of the free surface obtained from the measured water surface elevation. The no-slip boundary condition was prescribed at the wall boundaries and bridge piers. A range of flow rates for the Sacramento River was assigned at the inlet, as discussed in Section 6. Before conducting the LESs of the study reach, a short and straight reach with uniform channel geometry derived from the inlet cross-section of the study reach. This short channel was used to develop turbulent inlet boundary conditions at the inflow boundary of the study reach. Separate precursor runs were completed for each discharge used in this study. For each flow rate, when the precursor simulation reached a fully developed, turbulent flow state, the instantaneous velocity field at the outlet was stored for later use in the LESs of the study reach. No additional flow was assumed to enter the flow domain at the confluence with Sycamore Slough.

## 5 Uncertainty in model parameters

The hydrodynamic results of our LESs are presumed to be highly dependent on two crucial input parameters: (1) the inflow discharge and (2) the equivalent roughness height. As shown in Table 1, in the study reach, several different flow rates were measured using the ADCP. Using the mean value of 130.5 m3/s would be a logical choice for inflow discharge. However, one needs to determine how much uncertainty this mean discharge value will introduce in the LES results? To quantify this uncertainty, a range of nine different flow rates representing flow rates from 10 percent less than the mean discharge to 10 percent greater than the mean value was simulated. In addition, the equivalent roughness height, $k_s$, used in the wall model (Eqn. 7) could impact the computed velocity profiles within the river and possibly redistribute the velocities laterally, as well. As stated earlier, there is a lack of consensus among researchers for how to define $k_s$. For this study, several values were used to represent a wide range of potential values for $k_s$. Specifically, values for the equivalent roughness height ranging between $1.5d_{50} < k_s < 3d_{90}$ were utilized based on the disparity of values and relationships related to $k_s$ in literature (Camenen et al., 2006; Chanson, 2004; Ferguson, 2007; Hey, 1979; Sumer et al., 1996; van Rijn, 1984). Furthermore, it was assumed that the uncertainty associated with $k_s$ is uniformly distributed. The USGS sampled bed material from the river on November 1, 1977, at the USGS Gage No. 11391000 (*USGS 11391000 SACRAMENTO R A KNIGHTS LANDING CA Water Quality Data*, n.d.). Based on three representative sediment samples presented in Fig. 10, $d_{50} = 7.0\ mm$ and $d_{90} = 21.6\ mm$ making the equivalent roughness height to range between 10 mm and 65 mm. Therefore, for this study, values of $k_s$ equal to 10, 14, 20, 40 and 60 mm were used in the LESs. Due to the computational cost for running each LES, only 11 cases combining different discharges and equivalent roughness heights were run (see Table 2). To augment the "gaps" for other combinations of these two parameters, a PCE was used to estimate the flow field results of all possible combinations of parameter values for the inflow rate and equivalent roughness height.

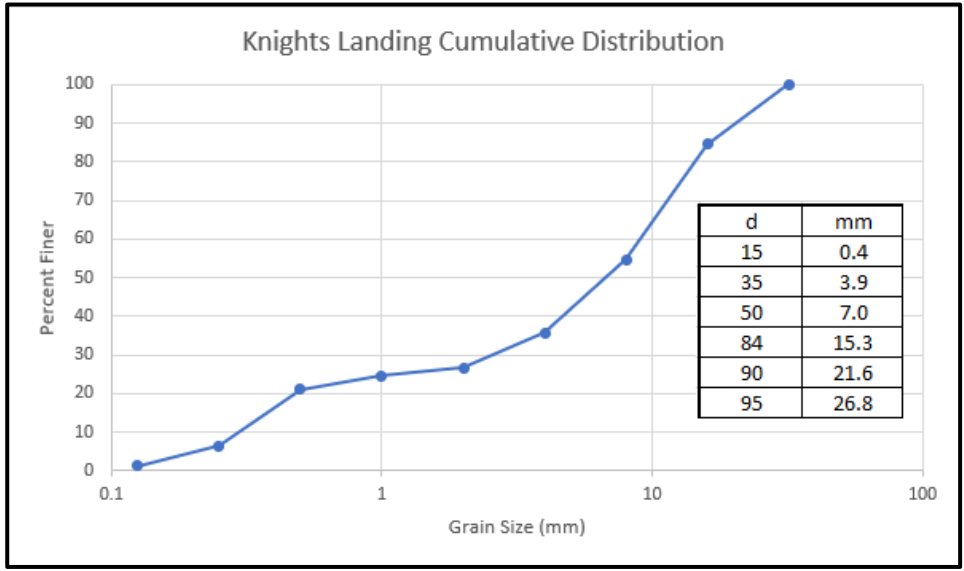

**Figure 10.** Cumulative size distribution for sediment samples of the Sacramento River at Knights Landing obtained from USGS Gage 11391000. In the insert table, d is the percent finer of the particle size in mm.

**Table 2.** Summary of LES scenarios. $k_s$ is the equivalent roughness height, $k_s^+$ is the roughness Reynolds number, $Q$ is dimensional flow rate, and the last column shows the percentage of flow relative to the mean flow rate (130.5 m³/s).

| Case Number | $k_s$ (mm) | $k_s^+$ | Mean $Q$ (m³/s) | Percent of Mean $Q$ |
|---|---|---|---|---|
| 1 | 10 | 150 | 130.5 | 100 |
| 2 | 14 | 210 | 130.5 | 100 |
| 3 | 20 | 300 | 130.5 | 100 |
| 4 | 20 | 300 | 140.3 | 107.5 |
| 5 | 40 | 600 | 130.5 | 100 |
| 6 | 60 | 900 | 130.5 | 100 |
| 7 | 60 | 900 | 120.7 | 92.5 |
| 8 | 40 | 600 | 137.0 | 105 |
| 9 | 40 | 600 | 124.0 | 95 |
| 10 | 20 | 300 | 120.7 | 92.5 |
| 11 | 60 | 900 | 140.3 | 107.5 |

The PCE is a general methodology that can be used in uncertainty quantification evaluations, representing the functional relationship between input parameters and the resulting output rather than only the statistical correlation between the parameter inputs and hydrodynamic outputs (Adams et al., 2020). Herein, non-intrusive PCE simulations are used as black boxes, and the calculation of the PCE coefficients for flow velocities is based on the set of LES cases (Eldred, 2009). Once the correct polynomial coefficients are determined, the PCE can approximate finite-dimensional series expansions of the flow velocities for any input variables. The chaos expansion for Response R can be approximated as (Eldred, 2009):

$$R \cong \sum_{j=0}^{P} \alpha_j \psi_j(\xi) \tag{8}$$

where $\alpha_j$ is a deterministic coefficient, $\psi_i$ is the ith multidimensional orthogonal polynomial, and $\xi$ is a vector of standardized random variables. Herein, the Askey family of orthogonal polynomials are employed for the PCE evaluation. More specifically, since the probability distributions for the inflow and $k_s$ are assumed to be uniformly distributed, the Legendre orthogonal polynomials are used for the PCE(Narayan & Xiu, 2012).

To assess the relative sensitivity caused by different variables in the uncertainty of the hydrodynamics results, the coefficients in PCEs can be analyzed using variance-based decomposition to extract values known as "Sobol" indices (Sudret, 2008). This analysis identifies the fraction that either individual variables or a specific combination of variables contribute to the total uncertainty in the response, Y. This effect is known as the main effect sensitivity index, and for a single variable, $x_i$, is represented by the Sobol Index value, $S_i$, which is defined as (Eldred et al., 1999):

$$S_i = \frac{Var_{x_i}[E(Y|x_i)]}{Var(Y)} \tag{9}$$

where $Var_{x_i}[E(Y|x_i)]$ represents the variance of the conditional expectation of Y for a given $x_i$ and $Var(Y)$ is the total variance.

Once the PCE is evaluated, the full range of inflow rates and equivalent roughness heights can be explored using the surrogate model by the MC sampling. The MC sampling repeatedly selects random values of the two parameters and evaluates the corresponding velocities from the PCE to define the confidence intervals from the statistical variance in the PCE responses. This process is repeated for each location in the study reach where confidence intervals are desired. Herein, we evaluate the depth-averaged velocities along 4 of the ADCP locations and at four vertical profiles, where ADCP measurements were taken. The Design Analysis Kit for Optimization and Terascale Applications (Dakota) software developed by Sandia National Laboratory was used to conduct the UQ analysis. Based on the 11 LES cases, Dakota first determined the PCE coefficients and Sobol indices for the points of interest in the study reach and then randomly sampled 10,000 combinations of parameter inputs using MC sampling to define 95 percent confidence intervals. Fig. 11 shows the workflow for the UQ analysis.

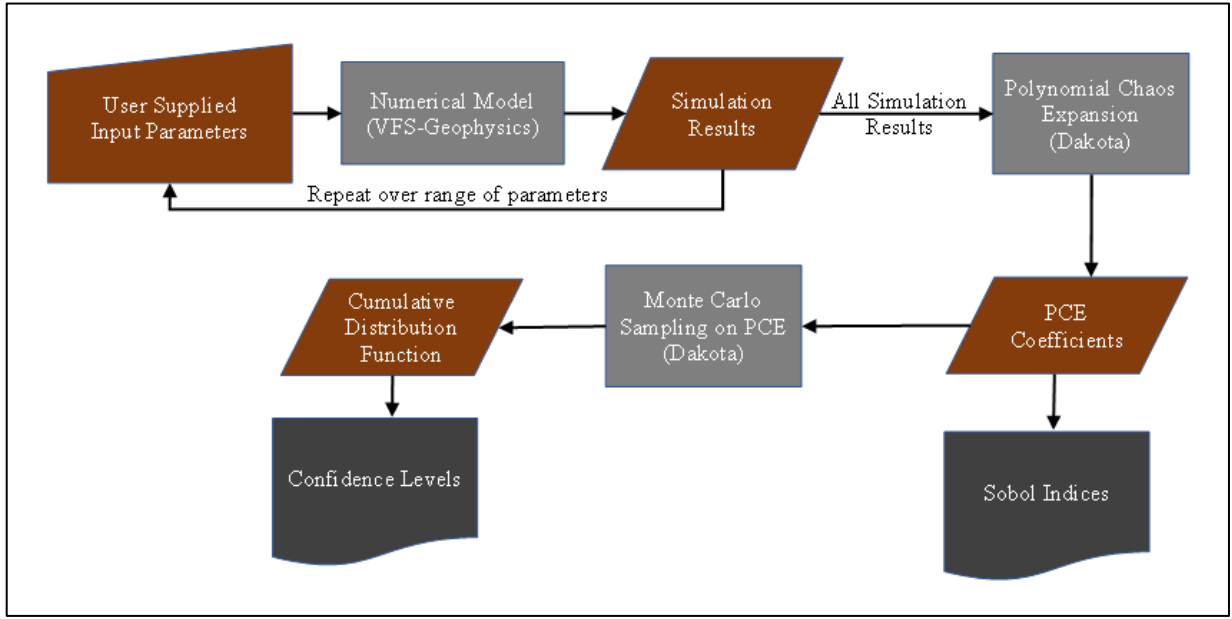

**Figure 11.** The flowchart illustrates the process of UQ analysis.

## 6 Results

Each of the 11 LES Cases listed in Table 2 ran until flow was well established through the reach, as shown in Fig. 12. This figure shows a large flow separation region occurring along the outer bend of the river near the confluence with Sycamore Slough. A smaller separation zone is also noted at the inner bend. Upstream of the bend, flow is relatively uniform, showing only velocity fluctuations corresponding to the bedforms on the left side of the channel. Due to the meander bend, however, the flow becomes noticeably less streamlined and exhibits continuous large-scale turbulence. After the kinetic energy of the flow stabilized, time-averaged results were computed until the mean statistics converged. Like the instantaneous flow, the small fluctuations in the time-averaged velocity magnitudes on the water surface reflect the presence of the channel bedforms throughout the channel and the flow separation along the right bank (Fig. 13 (a)). As seen in Fig. 13 (c), the sharp meander bend leads to large-scale redistribution of the velocity core in the water column. The strong secondary flow at the bend produces non-typical velocity profiles in vertical, which can be clearly seen in cross-section 3 of Fig. 13(c). The velocity magnitude at the water surface elevation is significantly lower than that of mid-depth near the outer bank. Trying to measure the discharge in this area will likely create inaccurate readings. For instance, if an ADCP is used, the flow near the water surface cannot be measured directly and is estimated using a logarithmic profile assumption which is not valid at this location. Similarly, when processing LSPIV data, the velocities at the water surface are typically reduced to estimate the depth-averaged value for use in the discharge calculation. As shown, reducing the surface velocity at this location would be inappropriate based on the computed results.

Due to the uncertainty in the inflow discharge, several LESs were conducted using identical equivalent roughness heights while varying only the flow discharge to assess the effect of flow discharge alone on the computed hydrodynamics. For instance, both Cases 4 and 10 simulated the flow field using $k_s = 20$ mm but varied the inflow discharge from 140.3 m3/s to 120.7 m3/s that are 107.5 percent and 92.5 percent of the mean flow, respectively. The time-averaged surface velocity magnitudes for these two cases are shown in Fig. 14(a) and (b). As expected, the higher flow rate in Case 4 shows noticeably higher velocity magnitudes at the water surface. However, when comparing the difference in velocity magnitudes in Fig. 14(c), the increase in velocity magnitude at the water surface scales nonuniformly concerning the 15 percent difference in discharges. Upstream of the bend along the inner bend of the river in the shallower parts of the river, flow velocity magnitudes increase by roughly 30 percent, whereas, along the outer bend, the velocity magnitudes increase by less than 10 percent. Even more surprisingly, the velocity field throughout the flow depth scales nonuniformly with the increased flow rate but shows highly variable increases throughout the study reach (Fig. 14 (d)). We argue that this could be attributed to the complex secondary flows and nonuniformity of bed bathymetry. The former is known to cause the lateral movement of the high-velocity core (Kang et al., 2011; Kang & Sotiropoulos, 2012), while the latter is known to produce heterogeneous turbulent flow structures in the water column (Khosronejad, Hansen, et al., 2016; Le et al., 2019).

Like the variability in velocity field throughout the river caused by changing the discharge, LES cases were also computed using different equivalent roughness heights while maintaining the same flow rate at the inlet. Fig. 15 shows the results from Case 1 and Case 6, which used $k_s$ values of 10 mm and 60 mm, respectively. As seen in Fig. 15 (a) and (b), the velocity magnitudes at the water surface appear to be quite similar; however, considering the difference in the two surface velocity magnitudes, one can see that increasing the bed roughness results in a decrease in the velocity magnitude ranging from 5 to 15 percent along the left bank (see Fig. 15(c)). Conversely, downstream from the bend along the right bank, increasing the bed roughness leads to an increase in surface velocities by 5 to 15 percent. Within the bend itself, these effects are even more evident, causing the same velocity changes to be even more significant. As seen in Fig. 15 (d) at the three cross-sections in different parts of the bend, the degree and location of the velocity magnitude changes are nuanced, with varying changes being induced by the roughness seen in other velocity structures in the cross-sections.

Typically, the velocity profile of a river flow follows a logarithmic or parabolic profile (Robert, 2003). Accordingly, the depth-averaged velocity magnitude is often approximated to be about 85 percent of the surface velocity magnitude (Buchanan et al., 1976). The surface velocity magnitude and depth-averaged velocity magnitudes are compared for Case 3 to investigate the 85 percent approximation using our simulation results in the study reach of the Sacramento River. This case corresponds to the inflow rate of 130.5 m3/s and the equivalent roughness height of $k_s = 20$ mm. As seen in Fig. 16 (a) and (b), as expected, the velocity magnitude is generally lower for the depth-averaged values. Closer examination of the percentage of the depth-averaged velocity magnitude relative to the surface velocity magnitude in Fig. 16(c), however, reveals that the percentage varies appreciably from the 85 percent estimate with blue regions being a higher percentage typically occurring along the inside of the bend and red regions being a lower percentage located principally along the outside of the bend. Based on the four cross-sections plotted in Fig. 16 (d), the range for the ratio varies considerably in the bend, i.e., from 70 percent to 95 percent. However, in cross-section B, extreme values that range from 50 percent to as much as 160 percent are found near the outside of the bend. Cross-section D is located near the outlet of the flow domain and shows the closest agreement with the 85 percent approximation while only deviating by about 5 percent higher and lower across most of the study reach, except near the banks where there is less agreement.

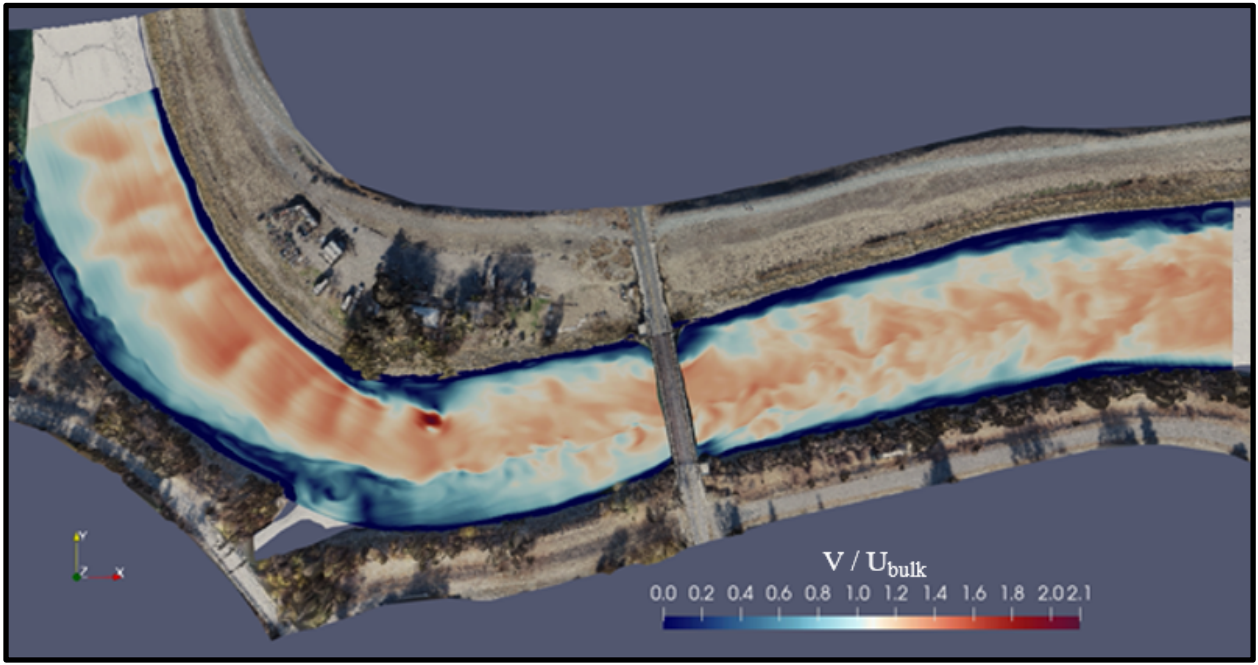

**Figure 12.** Contours of the instantaneous velocity magnitude ($V$) at the water surface. $V$ is normalized with the mean-flow velocity ($U_{bulk}$=0.565 m/s). The mean-flow depth is 3.3 m and is conveyed from left to right.

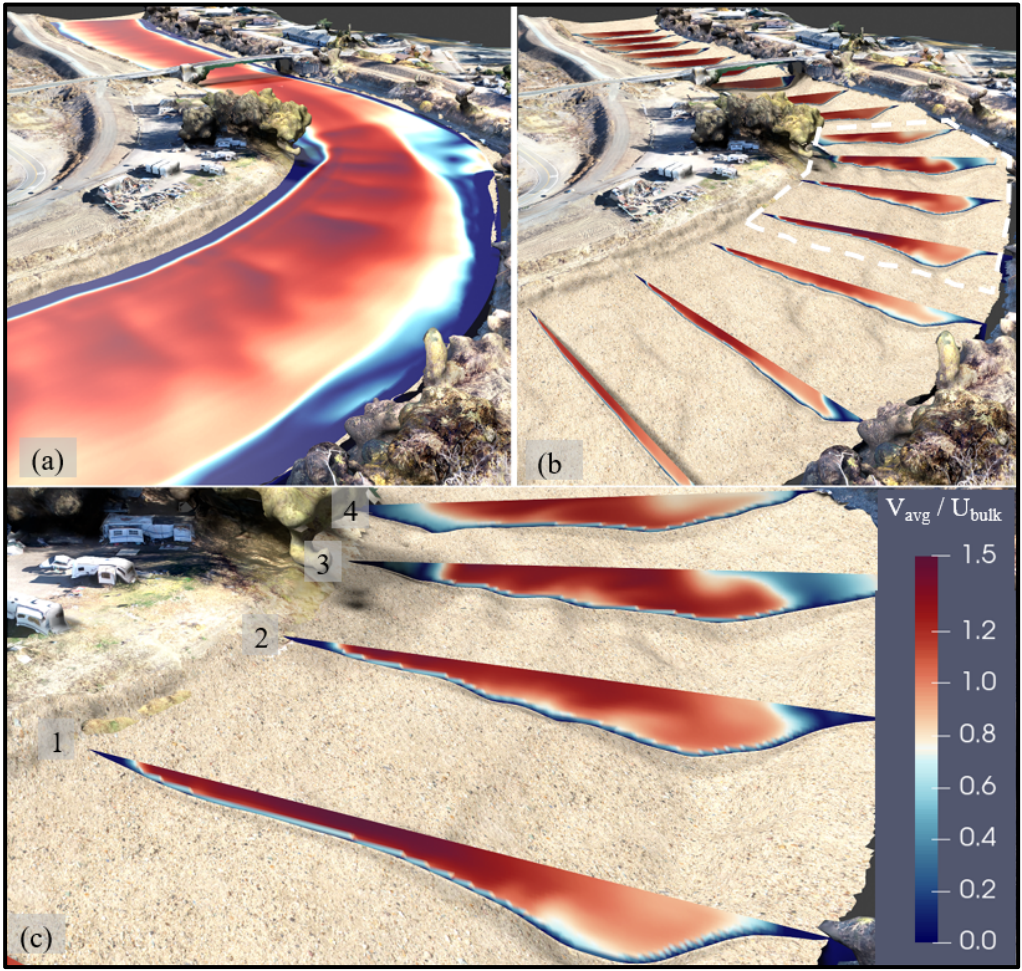

**Figure 13.** Contours of the time-averaged flow velocity ($V_{avg}$) normalized by the mean flow velocity ($U_{bulk}$=0.565 m/s). The free surface velocities are shown in (a), with flow moving from the lower left to the top. Slices are taken through the water column throughout the study reach and are presented in (b), with the white dashed area enlarged in (c).

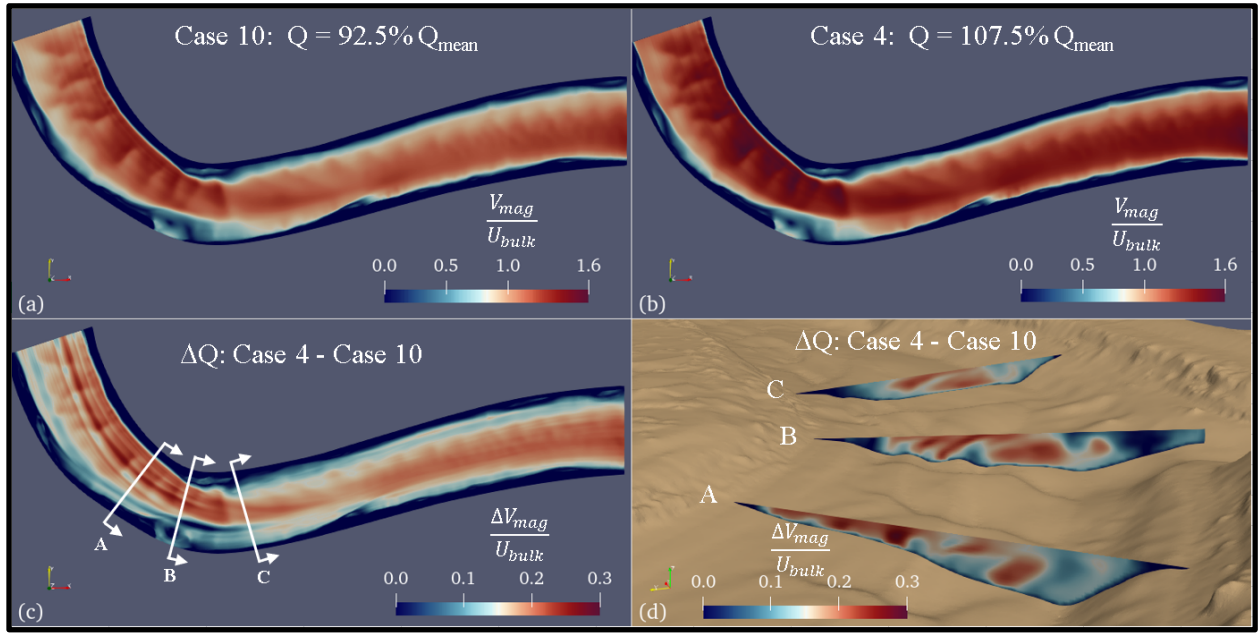

**Figure 14.** Comparison is shown for how different flow rates affect the velocity in the river. Velocity magnitudes normalized by the mean flow velocity ($U_{bulk}$) at the water surface are shown for flow rates of 92.5percent and 107.5 percent of the mean-flow rate (130.5 m$^3$/s) in (a) and (b), respectively. In (c), contours of the difference in surface velocity magnitudes for the two flow rates are shown. The three cross-sections drawn with white lines are presented enlarged in (d). Flow is from left to right.

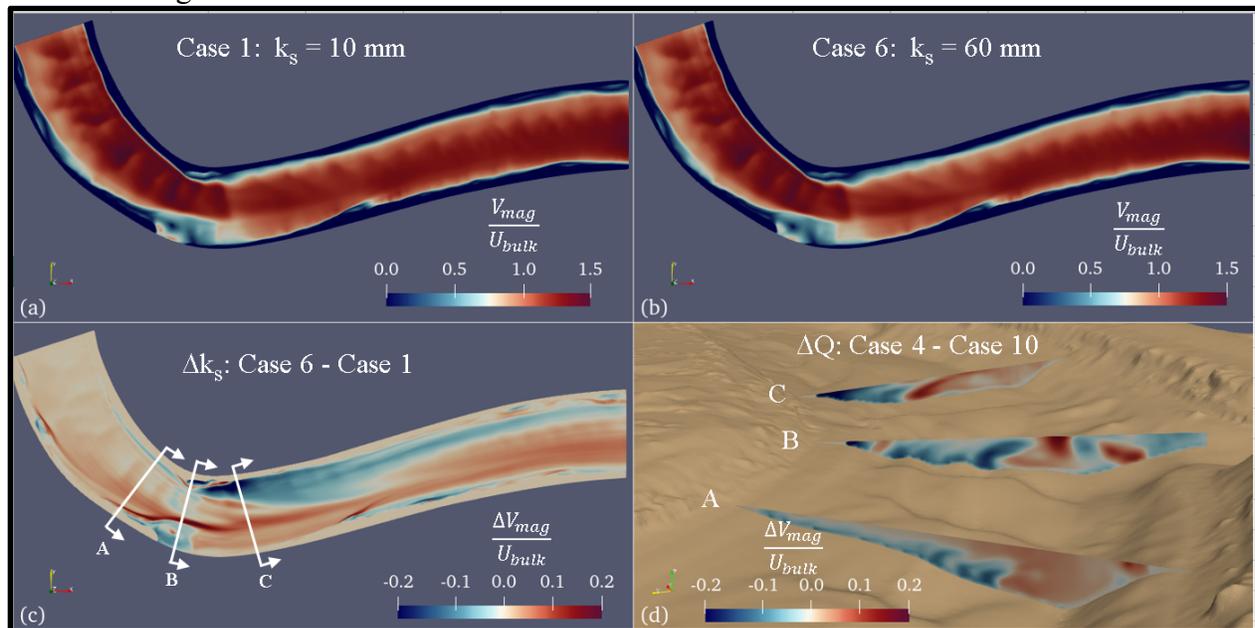

**Figure 15.** Comparison is shown for how different equivalent roughness heights affect the velocity in the river. Velocity magnitudes normalized by the mean flow velocity ($U_{bulk}$) at the water surface are shown for two values of $k_s$ (10 mm and 60 mm) in (a) and (b), respectively. In (c), contours of the difference in surface velocity magnitudes with two different $k_s$ values of 10 mm and 60 mm are shown. The three cross-sections drawn with white lines are enlarged in (d). Flow is from left to right.

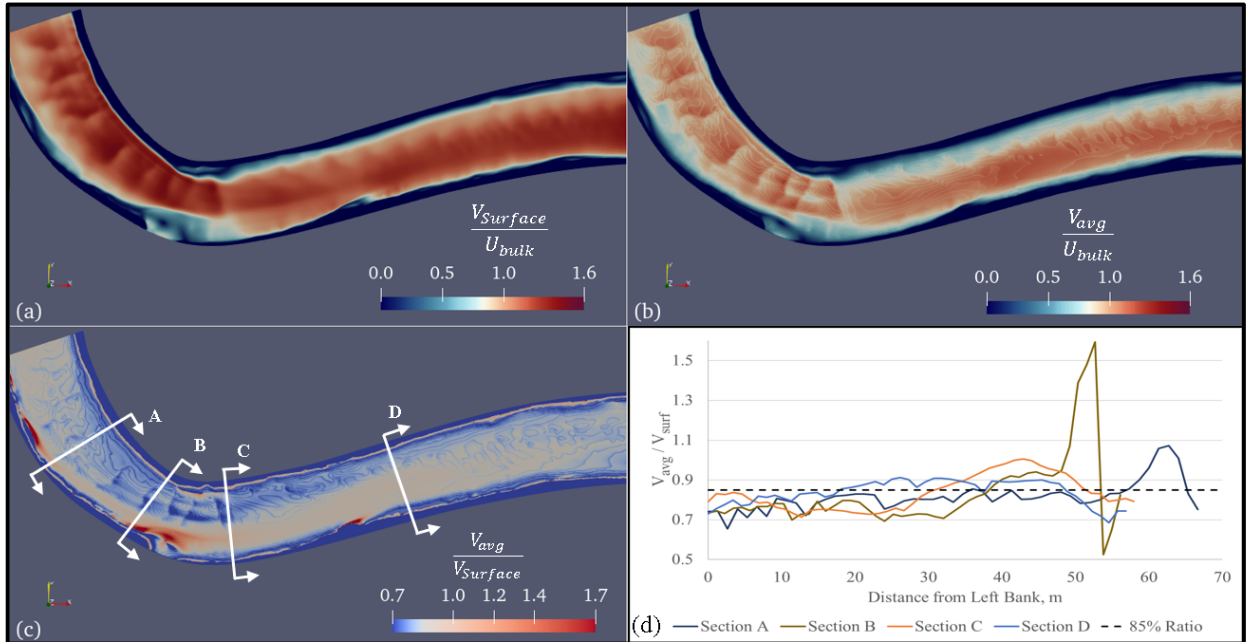

**Figure 16.** Contours of the LES-computed time-averaged velocity magnitudes normalized by the mean-flow velocity ($U_{bulk}$) at the water surface ($V_{surface}$) (a) and depth-averaged ($V_{avg}$) (b). (c) shows contours of $V_{avg}$ / $V_{surface}$. (d) plots the variation of $V_{avg}$ / $V_{surface}$ in the spanwise direction of the river along the cross-sections A to D, as shown in (c).

Thus far, we have demonstrated the strong yet somewhat unpredictable effect that both the inflow discharge and equivalent roughness height can have on the velocity field of the river flow in the study reach. Of principal interest in this study is the potential combined effect of these two uncertain parameters on the LES results, as expressed by confidence intervals. In other words, we seek to determine, with 95 percent confidence, the velocity magnitude of the river in the study reach, considering the uncertainty associated with the flow discharge and equivalent bed roughness. The computed velocities along selected cross-sections and velocity profile points were sampled using the 11 LES cases to determine the joint effect. Specifically, Locations 2 to 5 were selected along the mean cross-section location corresponding to the ADCP measurements, as shown in Fig. 5, where 16 equally spaced depth-averaged velocities were determined and used for the Uncertainty Quantification analysis. Similarly, the four-point locations (two along the cross-section at Location 2 and two along the cross-section at Location 5 (Fig. 7) were selected for samples of velocity profiles. Depending on flow depth, between 21 and 28 velocity magnitude values were extracted from the LES results at each point location and labeled as VP1 to VP4. These values were then used in the software program, Dakota, to determine the PCE coefficients and Sobol indices. Subsequently, the PCE results were utilized to determine the confidence intervals by randomly sampling 10,000 combinations of discharge and equivalent roughness heights combinations using MC simulations. The range of variables used in the MC simulations was as follows. Inflow discharge ranged uniformly within 10 percent of the mean discharge, and $k_s$ was varied uniformly between 10 mm and 65 mm. The results of these evaluations are presented in Figs. 17 to 24.

Considering the uncertainty in the depth-averaged velocity results for Location 2, which mainly resides farthest upstream in the study reach, Fig. 17(b) shows that the computed results from the LES correspond well with the recorded ADCP measurements. As seen, most of the ADCP measured data points fall within the 95 percent confidence limits. Considering that these measurements reflect the average of only a few instantaneous measurements, readings outside of the 95 percent confidence bands may be due to the error in the measurements. Conversely, since Location 2 is near the inlet of the flow domain, it is likely that the velocity field imposed at the inlet cross-plane (i.e., as the inlet boundary condition for the LES) does not accurately reflect the proper boundary conditions. This is mainly due to the complexities associated with the meandering bends upstream of the study reach, which were not considered.

Furthermore, the computed velocities falling outside the 95 percent confidence bands are located near the side banks, and these "outliers" tend to propagate downstream at the other locations. The magnitude of the computational errors, however, tends to diminish in the downstream direction. This trend would be expected because the LES results are gradually corrected for the initial error in velocity distribution at the inlet as flow is conveyed downstream. Quantifying the uncertainty caused by the distribution at the inlet is not possible without additional bathymetry data, and a more extended study reach. Regarding the relative contribution to the uncertainty, the Sobol indices in Fig. 17(c) show that the flow discharge almost entirely influences the level of uncertainty in the depth-averaged velocities. This is evident in the Sobol indices that uniformly exceed 0.9 throughout the study reach and, in all cross-sections, including depths ranging from 2 m to over 7 m. This uniform impact of the flow rate on the overall uncertainty could be better understood by employing a constant equivalent roughness height for all precursor simulations, which were conducted to obtain the inlet boundary condition for the LESs. Since Location 2 is nearest to the precursor flow, the full effect of varying the roughness height may have not fully impacted the hydrodynamics of the river until farther downstream.

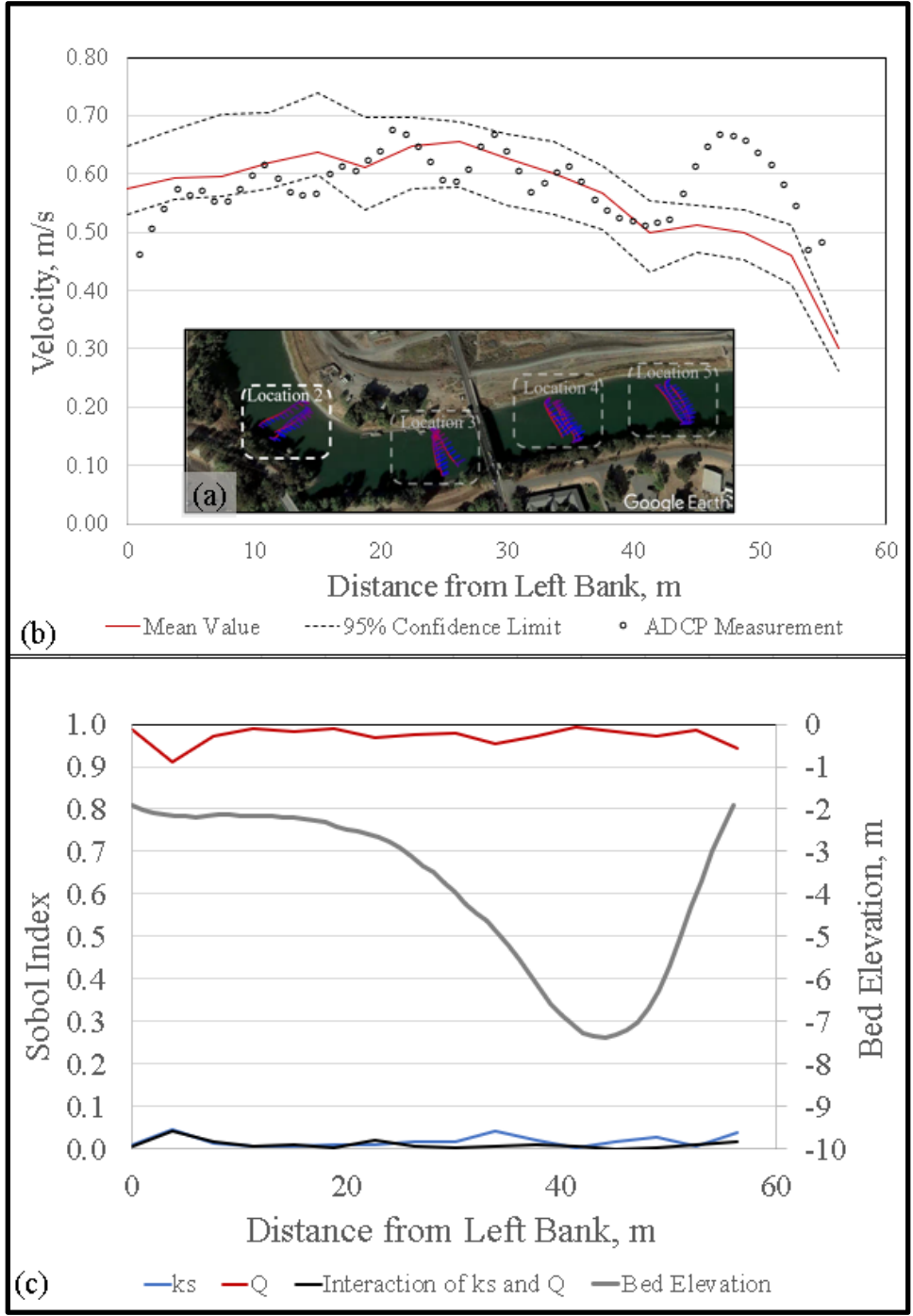

**Figure 17.** The depth-averaged velocity magnitudes normalized by $U_{bulk}$ in the spanwise direction of the river at Location 2. (a) shows the study reach and the cross-section at Location 2. In (b), the red solid-line shows the time-averaged LES results, circles are the instantaneously measured ADCP data and dashed lines mark the 95 percent confidence intervals. In (c), The Sobol indices for $Q$ (red line), $k_s$ (blue line), and the combined effect of $Q$ and $k_s$ (black line) are presented. The Sobol indices mark the relative influence of each parameter on the overall uncertainty of the LES results. For reference, the riverbed elevation (gray line) is plotted in (c).

Location 3 is immediately downstream from the river bend and the confluence with Sycamore Slough (Fig. 18(a)). Despite the strong secondary flow caused by the bend and the additional (although minor) flow entering at the confluence, the computed LES results are still in close agreement with the ADCP measured velocity field (Fig. 18(b)). As with Location 2, near each bank, the measured depth-averaged velocities are slightly outside of the 95 percent confidence bands. From the Sobol indices in Fig. 18(c), the uncertainty in the depth-averaged results is almost entirely due to the uncertainty in the discharge near the center of the channel; however, near the banks, a stronger influence of the bed roughness can be seen – particularly along the left bank where the uncertainty due to the equivalent roughness height is more pronounced.

At Locations 4, the cross-section is relatively uniform in shape (Fig. 19(c)) since it is located at a cross-over point of the river thalweg. Location 5, on the other hand, is located farther downstream of the study reach where the thalweg has shifted toward the left bank of the river (Fig. 20(c)). At both locations, the depth-averaged velocity measurements generally fall within the uncertainty of the LES results except near the side banks. However, at Location 5, which is farthest from the inlet precursor distribution, the measurements and numerical results are in closest agreement. As for the previous locations, the Sobol indices show that near the center of the river, the uncertainty in the computed results is strongly influenced by the uncertainty in the flow rate with values typically exceeding 90 percent. However, near the side banks, the uncertainty is mainly caused by the equivalent roughness height and ranges from 30 to 50 percent. It is worth noting that the confidence intervals near the side banks show less uncertainty than in the center of the river. Therefore, the increased Sobol index values do not necessarily indicate an absolute increase in the uncertainty caused by the bed roughness but may only be a more significant percentage of the total uncertainty in these regions.

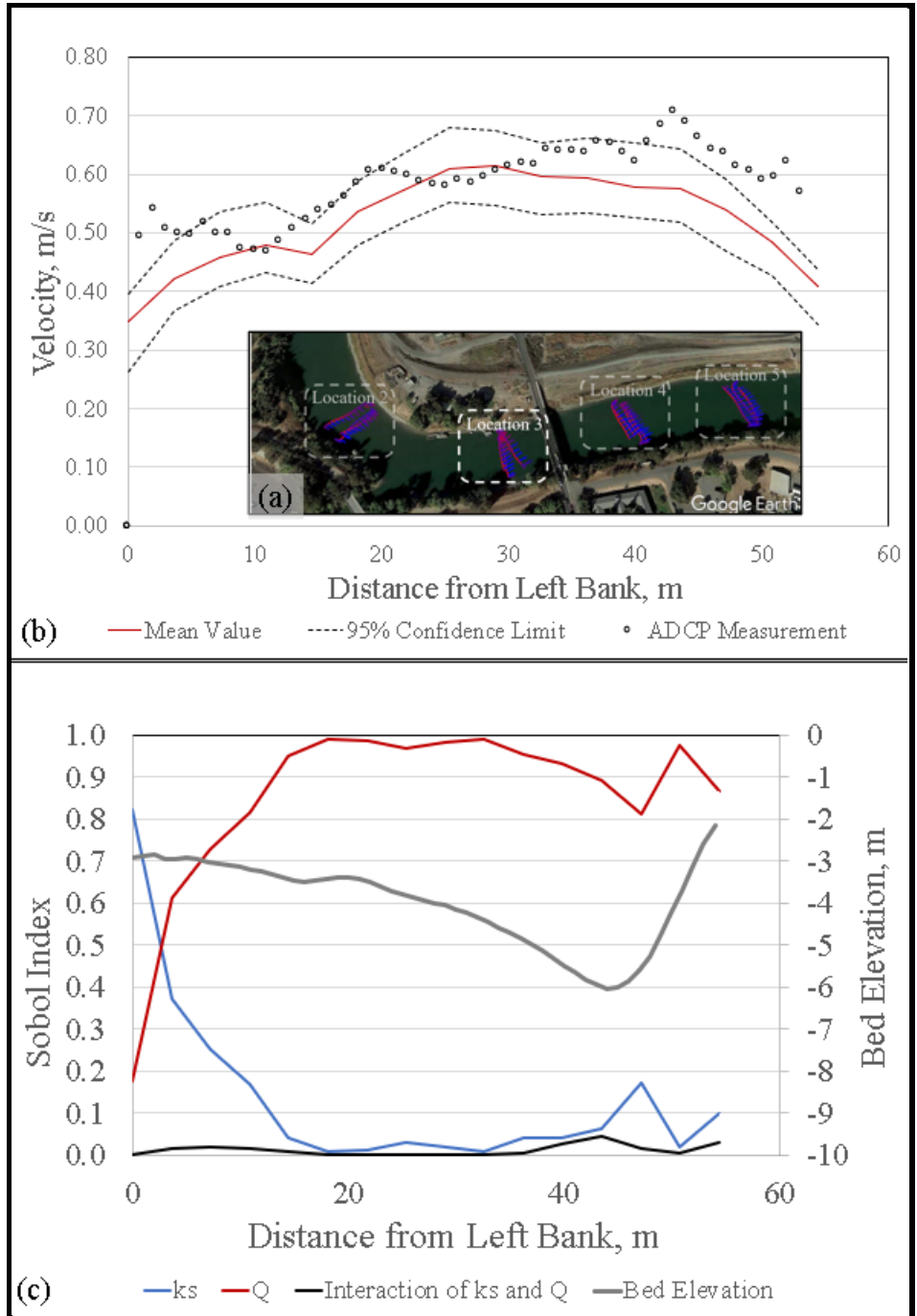

**Figure 18.** The depth-averaged velocity magnitudes normalized by $U_{bulk}$ in the spanwise direction of the river at Location 3. (a) shows the study reach and the cross-section at Location 3. In (b), the red solid-line shows the time-averaged LES results, circles are the instantaneously measured ADCP data and dashed lines mark the 95 percent confidence intervals. In (c), The Sobol indices for $Q$ (red line), $k_s$ (blue line), and the combined effect of $Q$ and $k_s$ (black line) are presented. The Sobol indices mark the relative influence of each parameter on the overall uncertainty of the LES results. For reference, the riverbed elevation (gray line) is plotted in (c).

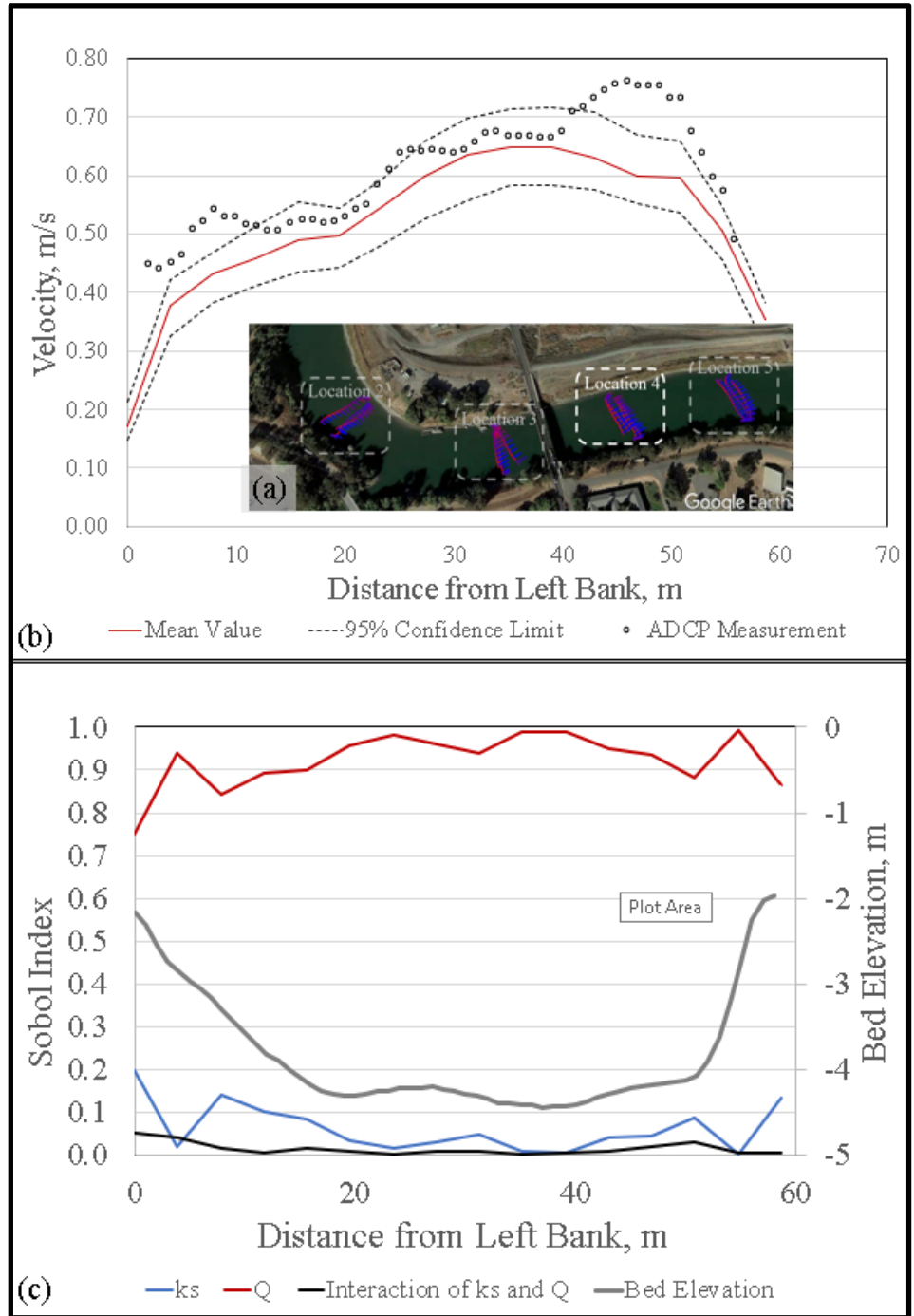

**Figure 19.** The depth-averaged velocity magnitudes normalized by $U_{bulk}$ in the spanwise direction of the river at Location 4. (a) shows the study reach and the cross-section at Location 4. In (b), the red solid-line shows the time-averaged LES results, circles are the instantaneously measured ADCP data and dashed lines mark the 95 percent confidence intervals. In (c), The Sobol indices for $Q$ (red line), $k_s$ (blue line), and the combined effect of $Q$ and $k_s$ (black line) are presented. The Sobol indices mark the relative influence of each parameter on the overall uncertainty of the LES results. For reference, the riverbed elevation (gray line) is plotted in (c).

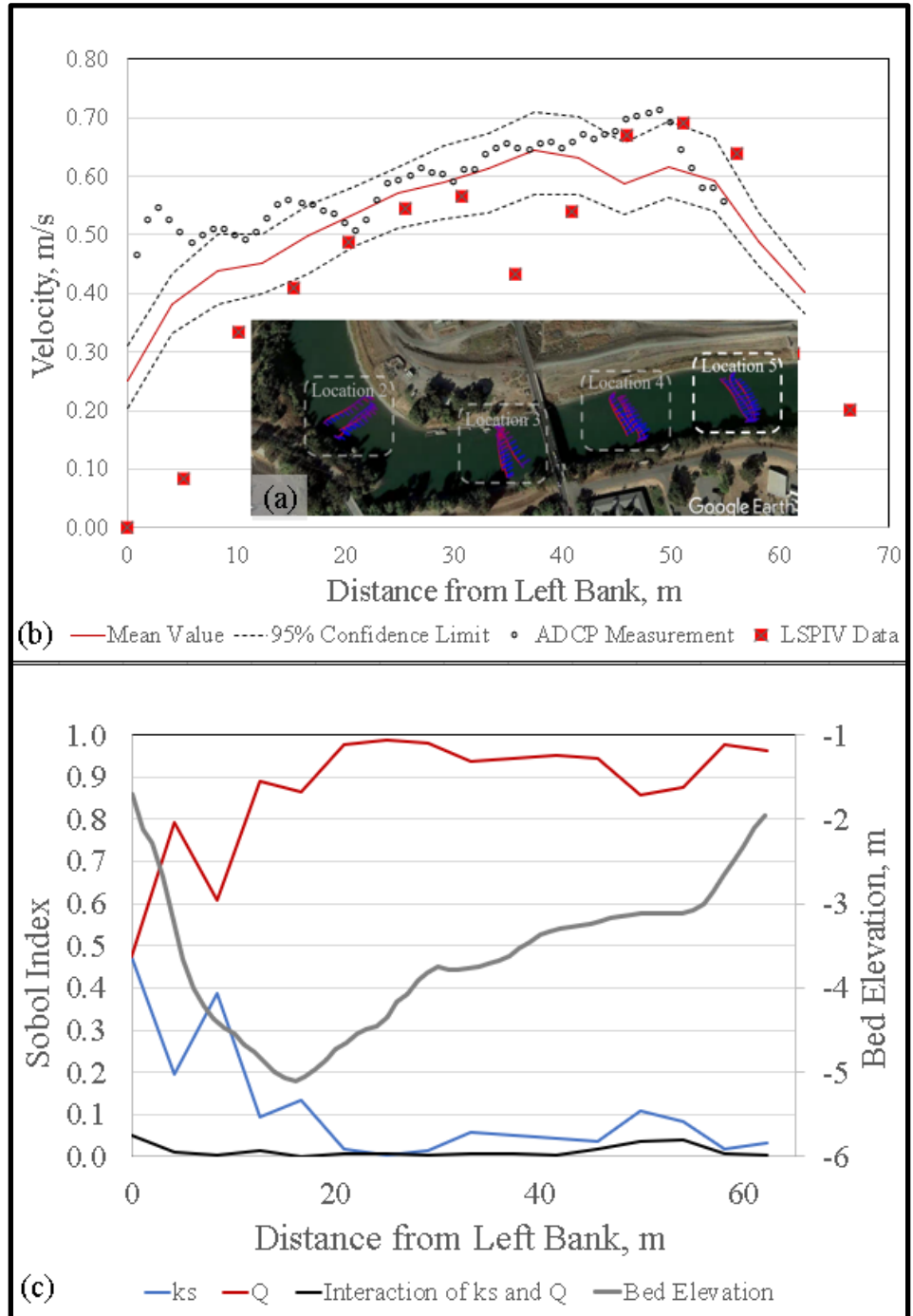

**Figure 20.** The depth-averaged velocity magnitudes normalized by $U_{bulk}$ in the spanwise direction of the river at Location 5. (a) shows the study reach and the cross-section at Location 5. In (b), the red solid-line shows the time-averaged LES results, circles are the instantaneously measured ADCP data and dashed lines mark the 95 percent confidence intervals. In (c), The Sobol indices for $Q$ (red line), $k_s$ (blue line), and the combined effect of $Q$ and $k_s$ (black line) are presented. The Sobol indices mark the relative influence of each parameter on the overall uncertainty of the LES results. For reference, the riverbed elevation (gray line) is plotted in (c).

The velocity profiles in the water column were also analyzed using the same UQ methodology to define the confidence bands and Sobol indices of the LES results at 4 locations within the study reach. Locations VP1 and VP2 are along the outer bank at the entrance to the meander bend on the same cross-section at Location 2 discussed above. For these two computed velocity profiles, most of the measured velocity data does not fall within the 95 percent confidence intervals. We conjecture that this may be primarily due to the measured values, which reflect only instantaneous velocity readings. While the depth-averaged ADCP readings in the spanwise directions were also computed by averaging only a few instantaneous readings, the depth-averaging process itself helps smooth out the instantaneous fluctuations by averaging the ensemble of values over the entire water column. However, the velocity profiles in the vertical direction do not have this same advantage and show only the average of 3 or 4 instantaneous readings. To better assess this error, one would need to make stationary measurements over several minutes to account for the various turbulent structures passing through a given point. We note that such stationary measurements in the field are extremely difficult. Locations VP3 and VP4 fall along the cross-section at Location 5, with VP3 and VP4 being located near the left and right banks, respectively (Fig. 23 (a) and (b)). At these two locations, the ADCP measurement data agree more closely with the LES results. This improved agreement may be due to the fact that these points are far away from the inlet and strong secondary effects due to the bend. For each case, the Sobol indices for $k_s$ increases appreciable from less than 5 percent near mid-depth to between 15 to 40 percent near the bed. This increase influence of equivalent roughness height on the uncertainty near the bed seems reasonable since the value for $k_s$ is directly applied to the wall model to reconstruct the velocity field of the computational nodes near the bed. However, it is also noted that the confidence limits near the riverbed are very small for each of the four vertical profiles indicating the uncertainty associated with $Q$ and $k_s$ are primarily experienced away from the bed. In this light, the influence of $k_s$ on the uncertainty should be recognized as higher relative to $Q$ and not as causing greater uncertainty in an absolute sense.

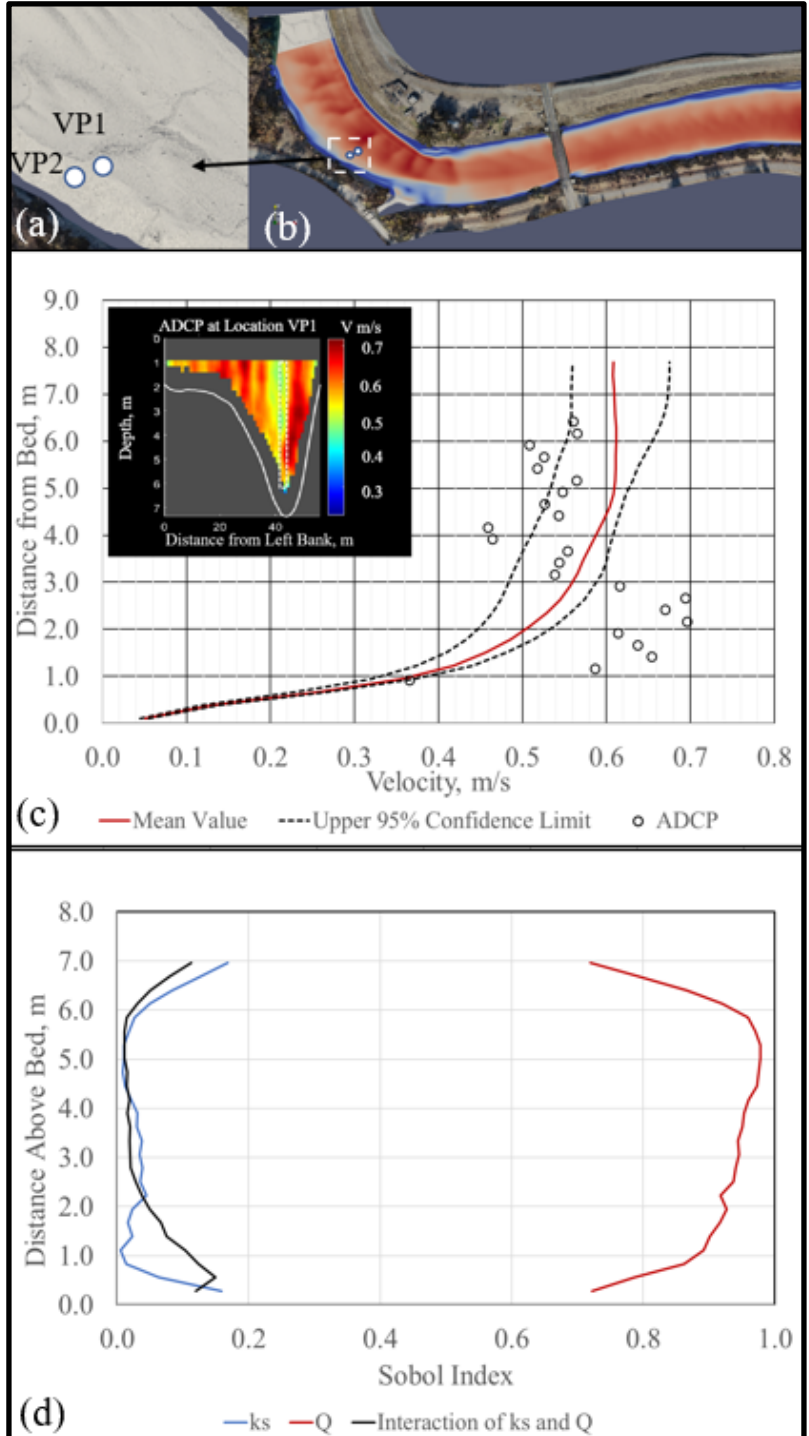

**Figure 21.** The vertical profile of velocity magnitude normalized by $U_{bulk}$ at point VP1. (a) and (b) illustrate the location of the VP1 over the riverbed. In (b), the red line shows the time-averaged LES results, dashed black lines mark the 95 percent confidence bands, and circles represent the instantaneous ADCP data. (d) depicts the Sobol indices for $Q, k_s$, and combined effect of $Q$ and $k_s$, showing each parameter's relative influence on the overall uncertainty.

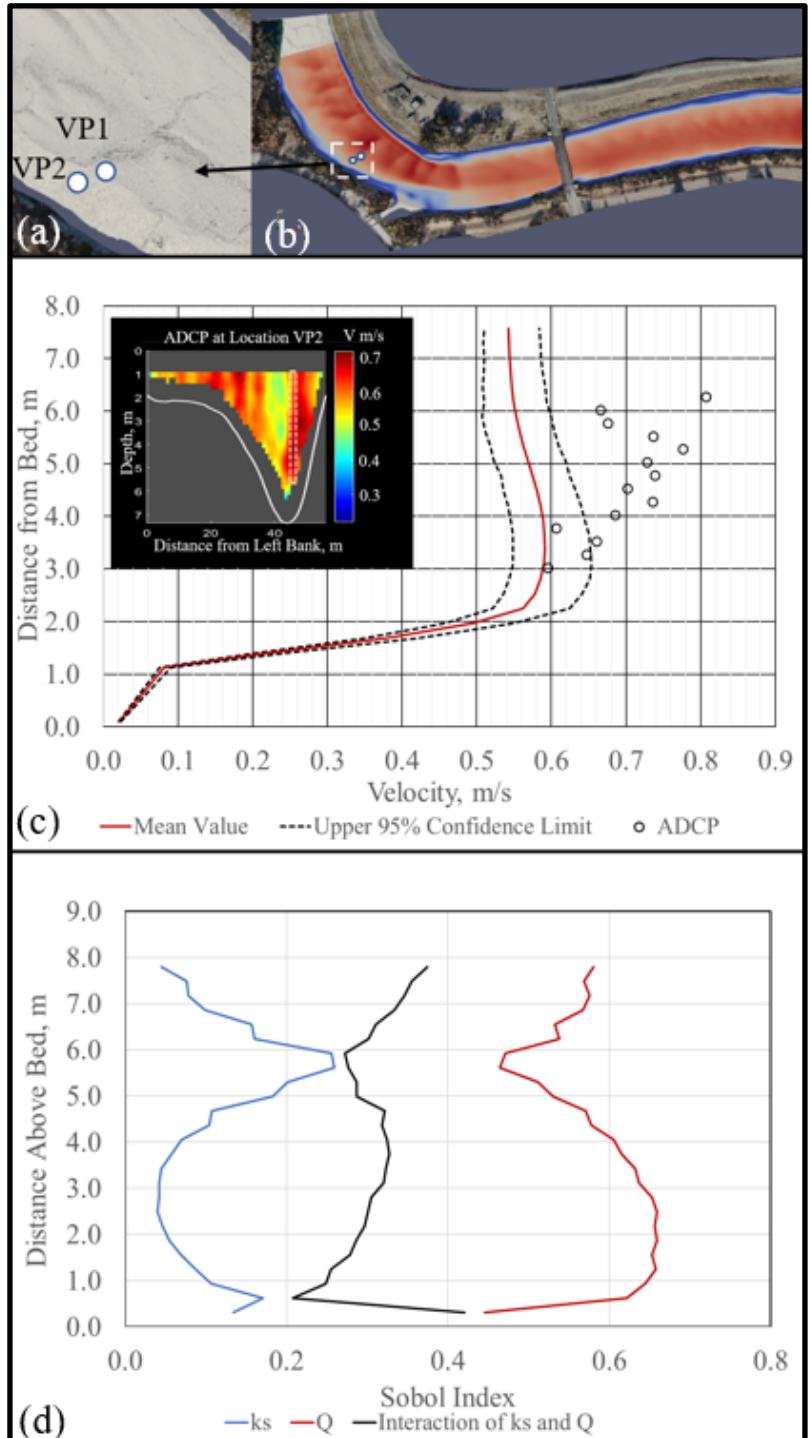

**Figure 22.** The vertical profile of velocity magnitude normalized by $U_{bulk}$ at point VP2. (a) and (b) illustrate the location of the VP1 over the riverbed. In (b), the red line shows the time-averaged LES results, dashed black lines mark the 95 percent confidence bands, and circles represent the instantaneous ADCP data. (d) depicts the Sobol indices for $Q$, $k_s$, and combined effect of $Q$ and $k_s$, showing each parameter's relative influence on the overall uncertainty.

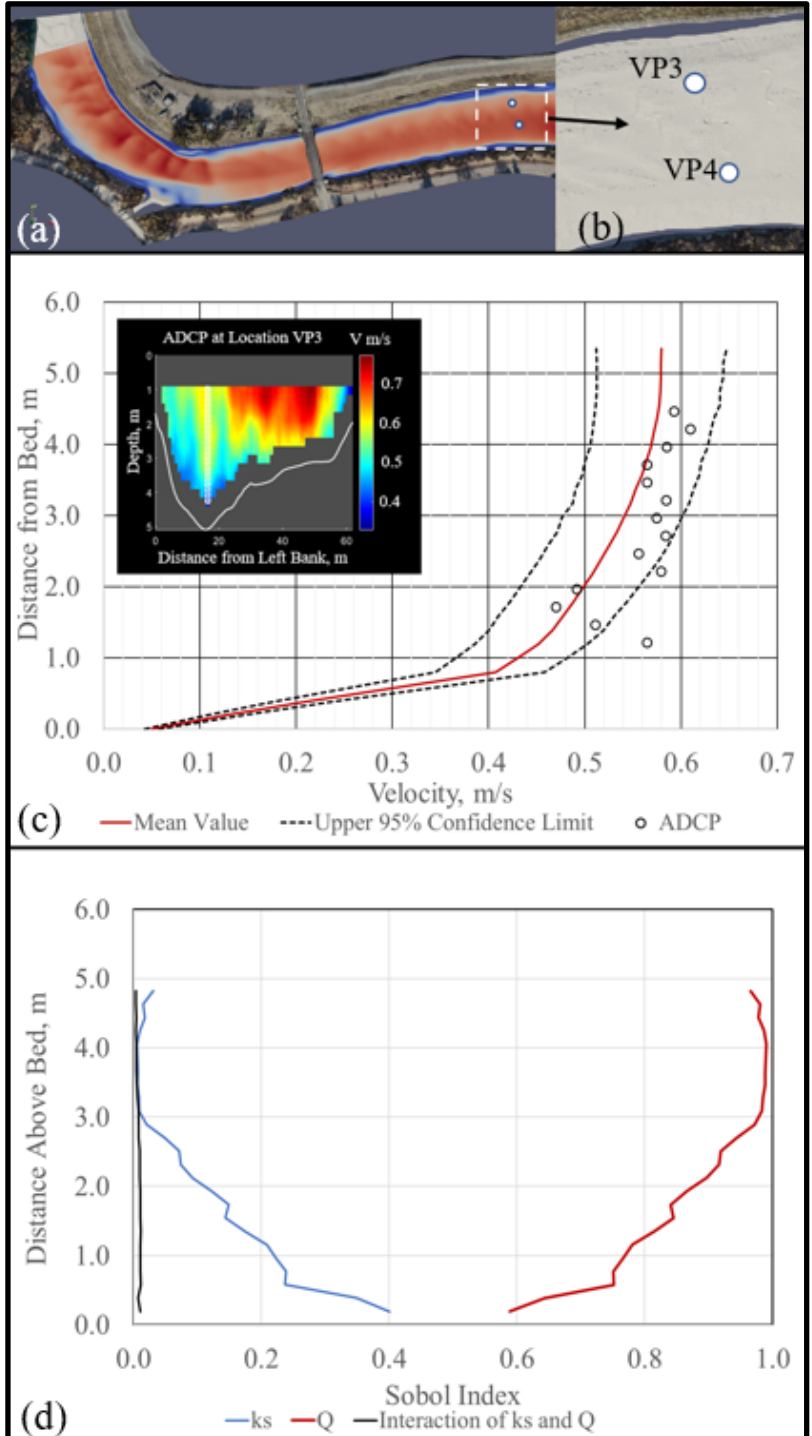

**Figure 23.** The vertical profile of velocity magnitude normalized by $U_{bulk}$ at point VP3. (a) and (b) illustrate the location of the VP1 over the riverbed. In (b), the red line shows the time-averaged LES results, dashed black lines mark the 95 percent confidence bands, and circles represent the instantaneous ADCP data. (d) depicts the Sobol indices for $Q, k_s$, and combined effect of $Q$ and $k_s$, showing each parameter's relative influence on the overall uncertainty.

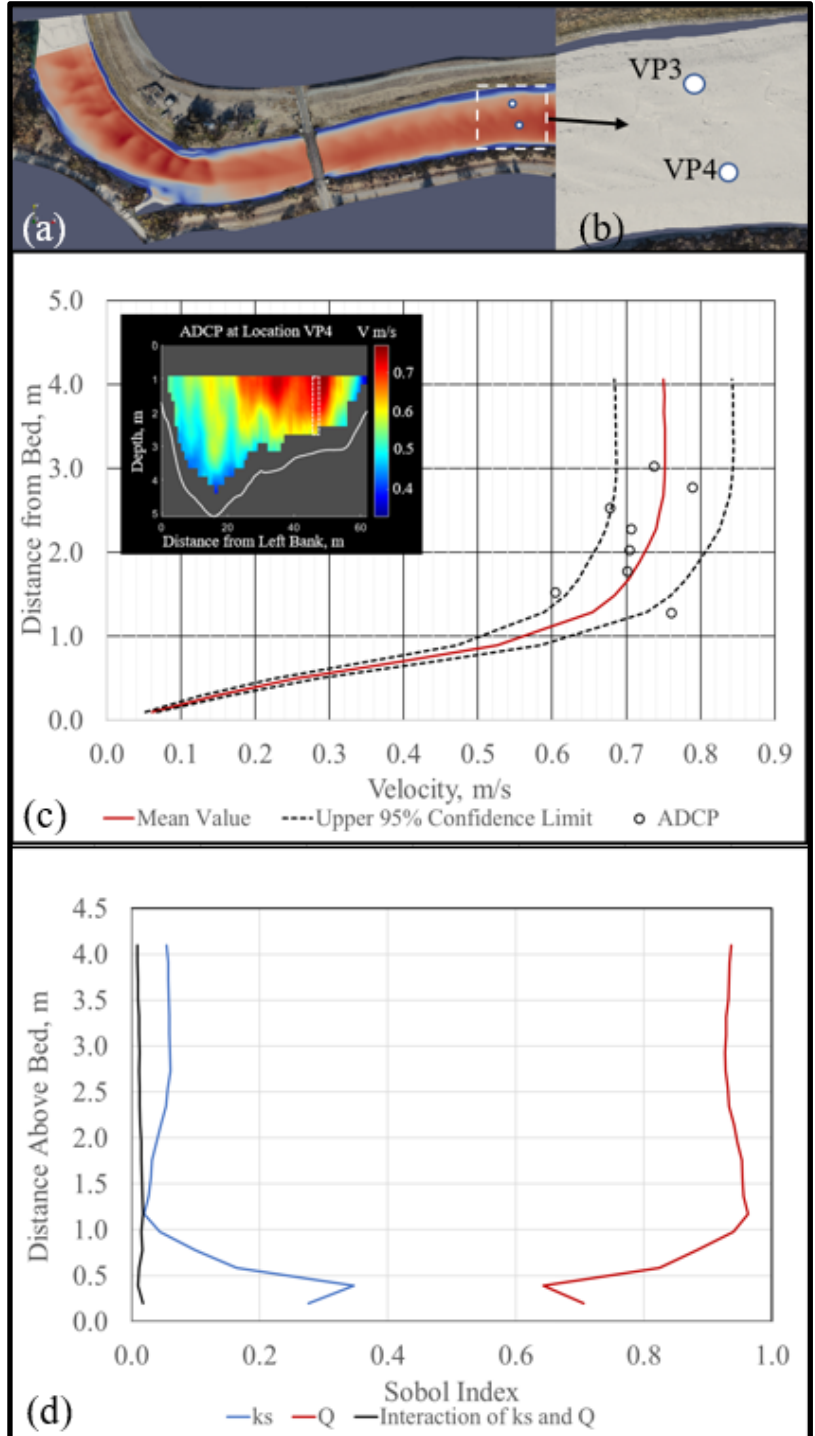

**Figure 24.** The vertical profile of velocity magnitude normalized by Ubulk at point VP4. (a) and (b) illustrate the location of the VP1 over the riverbed. In (b), the red line shows the time-averaged LES results, dashed black lines mark the 95 percent confidence bands, and circles represent the instantaneous ADCP data. (d) depicts the Sobol indices for $Q, k_s$, and combined effect of $Q$ and $k_s$, showing each parameter's relative influence on the overall uncertainty.

# 7 Conclusions

High-fidelity numerical modeling of turbulent flows in riverine systems with intricate bathymetry is generally based upon field measurements and modeling parameters – both of which have inherent uncertainty. In this study, we evaluated the propagation of uncertainties, due to the variations in the inflow discharge and roughness parameters, through LES computed hydrodynamics of the Sacramento River. When considered independently, both parameters redistributed the flow laterally in the river and modified the velocity profiles in nonuniform ways. The presence of the meander bend in the study reach and associated secondary flows created nonintuitive changes in the flow structure, making the relationship between depth-averaged velocity magnitudes and the surface velocities highly variable. However, farther downstream from the bend, the depth-averaged flow from the LES results seemed to be typically between 80 and 90 percent of the velocity magnitude at the free surface. This conclusion demonstrates the value of collecting field velocity data at locations in the river away from meander bends to obtain more predictable flow data.

The UQ analysis of the joint effect of equivalent roughness height and inflow discharge uncertainty provided 95 percent confidence bands of the depth-averaged velocities at selected locations which were generally validated by the ADCP measurements. The Sobol indices analysis showed that uncertainty in the flow discharge at the inlet contributed to most (over 90 percent) of the uncertainty of the computed depth-averaged velocities. However, at regions near the side banks, the equivalent roughness height contributed more influence ranging from 10 to 80 percent. This variability was associated with the location with the higher percentage of uncertainty attributed to the bed roughness, i.e., near the meander bend. Moreover, the uncertainty of the vertical velocity profiles near the bed was low and mostly was dominated by the equivalent roughness height. Away from the riverbed, the Sobol indices indicate that inflow discharge contributed almost exclusively to the uncertainty in the velocity profile.

While the results of this study apply to the specific flow conditions and river reach on the Sacramento River, they illustrate the importance of quantifying the uncertainty in both the model parameters and the resulting hydrodynamics when conducting numerical modeling of natural river flows. A proper UQ assessment of the model can provide helpful confidence levels for the computed results and, thus, can add valuable context for interpreting and applying the model to real-world applications.

## Acknowledgments

The LES code (DOI: 10.5281/zenodo.4677354) is available on Zenodo repository. This work was supported by the National Science Foundation (grant EAR-0120914). Computational resources were provided by the College of Engineering and Applied Science at Stony Brook University. The field campaign to measure bathymetric survey, conduct aerial survey, and collect ADCP data were supported by the California Department of Transportation.